\documentclass[preprint]{aastex}
\usepackage{graphicx}

\begin{document}

\title{MASS OUTFLOW AND CHROMOSPHERIC ACTIVITY OF RED GIANT STARS IN GLOBULAR CLUSTERS II. M13 AND M92}

\author{Sz. M{\'e}sz{\'a}ros\altaffilmark{1,2,3}, A. K. Dupree\altaffilmark{1,4} and T. Szalai\altaffilmark{2,5}}

\altaffiltext{1}{Harvard-Smithsonian Center for Astrophysics, Cambridge, MA 02138}
\altaffiltext{2}{Department of Optics and Quantum Electronics, University of Szeged, 6701 Szeged, Hungary}
\altaffiltext{3}{e-mail address: meszi@cfa.harvard.edu}
\altaffiltext{4}{e-mail address: dupree@cfa.harvard.edu}
\altaffiltext{5}{e-mail address: szaszi@titan.physx.u-szeged.hu}

\begin{abstract}

High resolution spectra of 123 red giant stars in the globular cluster M13 and 64 red giant stars in M92 
were obtained with Hectochelle at the MMT telescope. Emission and line asymmetries in
H$\alpha$, and \ion{Ca}{2}~K are identified, characterizing motions in the extended atmospheres and seeking
differences attributable to metallicity in these clusters and M15. On the red giant branch, emission in 
H$\alpha$ generally appears in stars with  $T_{eff} \la 4500$ K and log L/L$_\sun \ga$ 2.75. Fainter stars 
showing emission are asymptotic giant branch (AGB) stars or perhaps binary stars. The line-bisector for 
H$\alpha$ reveals the onset of chromospheric expansion in stars more luminous than log~$(L/L_{\odot})\sim2.5$ 
in all clusters, and this outflow velocity increases with stellar luminosity. However, the coolest giants 
in the metal-rich M13 show greatly reduced outflow in H$\alpha$ most probably due to decreased T$_{eff}$ 
and changing atmospheric structure. The \ion{Ca}{2}~K$_3$ outflow velocities are larger than  shown by 
H$\alpha$ at the same luminosity and signal accelerating outflows in the chromospheres. 
Stars clearly on the AGB show faster chromospheric outflows in H$\alpha$ than RGB objects. While the H$\alpha$
velocities on the RGB are similar for all metallicities, the AGB stars in the metal-poor M15 
and M92 have higher outflow velocities than in the metal-rich M13. Comparison of these chromospheric line
profiles in the paired metal-poor clusters, M15 and M92 shows remarkable
similarities in the presence of emission and dynamical signatures, and does not reveal a source of the
`second-parameter' effect.
\end{abstract}


\section{Introduction}

The well$-$known second-parameter problem in globular clusters \citep{sandage04}, 
in which a parameter other than metallicity affects the morphology of the horizontal branch, remains unresolved. 
Metallicity, as first noted by \citet{sandage03}, remains the principal parameter, but pairs of clusters,
with the same metallicity, display quite different horizontal branch morphologies thus challenging the canonical 
models of stellar evolution and leading to the need for a `second-parameter'.
Cluster ages have been  examined in many studies \citep{searle01, lee02, stetson01, lee01, sarajedini01, sarajedini02} 
and in addition, many other suggestions for the `second-parameter(s)' have been proposed,
including: total cluster mass; stellar environment (and possibly free-floating
planets); primordial He abundance; post-mixing surface
helium abundance; CNO abundance; stellar rotation; and mass loss \citep{catelan01, catelan02, sills01, soker01, 
sweigart01, buonanno01, peterson03, buonanno02, recio-blanco01}. Many authors \citep{vandenberg01, lee02, 
catelan01} have proposed that the second-parameter problem cannot be explained by only one second parameter. 
Various studies have attempted to limit the number of second parameters.

An example of paired second-parameter clusters is M15 and M92 ([Fe/H]=$-$2.26 and $-$2.28 respectively). 
Although the metallicities of these two clusters are the same \citep{sneden01}, their horizontal branches (HBs)  
differ \citep{buonanno03}. M92 has a brighter (by about one magnitude) and redder 
blue HB extension than M15. The color magnitude diagrams (CMDs)
of this pair were examined in detail by \citet{cho01}. They found that the difference in the HB morphology 
between the two is probably not a result of deep mixing in their red giant branch (RGB) sequences,
because no significant `extra stars' were found in their observed RGB luminosity functions compared to the 
theoretical RGB luminosity functions. \citet{sneden01} found that Si, Ca, Ti, and Na abundance ratios of the red 
giants are nearly the same in both clusters, only the [Ba/Ca] ratio shows a large scatter and the mean value in 
M15 is twice that found in M92. These studies eliminate deep mixing and subtle abundance variations as possible 
second parameters, while mass loss is examined in this paper. Detailed observations of red giant stars in M15 
are contained in \citet{meszaros01}, but the comparison between M15 and M92 is described here. 

M13 ([Fe/H]=$-1.54$) is one of the most studied second-parameter globular clusters. M13 and M3 are almost 
identical in most respects (metallicity, age, chemical composition), but there are dramatic differences in both 
the HB and blue straggler populations. Analysis of both clusters' CMDs \citep{ferraro01} with the Hubble Space 
Telescope revealed that neither age nor cluster density, two popular second-parameter candidates, is likely to be 
responsible for the differences in these clusters. From the analysis of high-resolution, high signal-to-noise 
ratio spectra of six RGB stars in M3 and three in M13, \citet{cavallo01} found that the [Al/Fe] and [Na/Fe] 
abundances increase toward the tip of the RGB. They concluded that the data for both clusters are consistent with 
deep mixing as a second parameter. Later, \citet{johnson01}, from medium$-$resolution spectra of more than 200 
stars in M3 and M13, concurred that deep mixing is the best candidate for second parameter in this pair of 
clusters. \citet{caloi01} also examined the second-parameter problem in M3 and M13 in detail and proposed that the 
overall difference between M3 and M13 CMD morphologies is due to the different helium content. Since M13 does not 
have a red clump in its horizontal branch they suggested that it represents an extreme case of self-enrichment of 
helium, which might come from the massive asymptotic giant branch stars (AGB) in the 
first $\approx$100 Myr of the cluster life.

A multivariate study of the CMDs for 54 globular clusters was carried out by \citet{recio-blanco01} from Hubble 
Space Telescope photometry to quantify the parameter dependencies of HB morphology. They found that the total 
cluster luminosity (therefore the total mass) has the largest impact on the HB morphology, and as \citet{caloi01} 
speculated, there may be enrichment of helium from an earlier population of stars. \citet{dantona01} modeled the 
evolution of globular cluster stars and showed that different choices of mass$-$loss rate affect the distribution 
of stars on the HB.

In this paper we discuss the characteristics of H$\alpha$ and \ion{Ca}{2}~K emission in M13 ([Fe/H]=$-$1.54) and 
M92 ([Fe/H]=$-$2.28). We compare our results
with the previously observed, metal-poor cluster M15 ([Fe/H]=$-$2.26) \citep{meszaros01} and the metal-rich cluster 
NGC~2808 ([Fe/H]=$-$1.15) \citep{cacciari01}. Detailed study of these four clusters allows us to examine a possible 
dependence between the average cluster metallicity and characteristics of H$\alpha$ and \ion{Ca}{2}~K emission, and 
diagnostics of mass outflow. Observations with 
the same instrument of the second-parameter pair M15 and M92 offer a good comparison to examine mass loss as a 
possible second parameter. 

\section{Observations and Data Reduction}

The Hectochelle on the MMT \citep{szentgyorgyi01} contains 240 fibers that can be placed $\sim$ 2 arcsec apart 
on the sky across the field of view which spans 1 degree. The diameter of one fiber on the sky is 1.6 arcsec. 
The apparent diameter of M13 is $\sim$15 arc minutes, and about 60$-$70 red giants in the globular cluster could 
be measured with each fiber configuration. The apparent diameter of M92 is smaller, and only 30$-$40 
stars could be measured  with one configuration. Two separate input fiber configurations for different stars were 
made for each cluster. A total of 123 different red giant stars in M13 and 64 red giants in M92 were observed in 
2006 March and 2006 May.

Targets with a high probability ($>95\%$) of membership were chosen from the catalog of \citet{cudworth01} for M13 
and from \citet{cudworth02} for M92. Smooth coverage of the RGB and AGB could be achieved within 
the constraint of the fiber placement on the sky. The CMD of the observed cluster members can be seen in Figure 1 
and the target stars are listed in Table 1 (M13) and Table 2 (M92). Coordinates of the stars were taken from the 
2MASS catalog \citep{skrutskie01} and used to position the fibers. Many fibers ($\sim$ 150$-$200) were placed on 
blank regions of the sky in order to measure the sky background in detail. 
These sky fibers were equally distributed in the Hectochelle field of view to cover a large area around the 
clusters. Hectochelle is a single-order instrument and three orders were selected for observation with 
order-separating filters: OB25 (H$\alpha$, region used for analysis $\lambda \lambda \ 6475-6630$), Ca41 
(\ion{Ca}{2}~H$\&$K, region used for analysis $\lambda \lambda \ 3910-3990$), and RV31 
(region used for analysis $\lambda \lambda \ 5150-5300$). Bias and quartz lamps for the flat correction were 
obtained during the afternoon each day. Exposures with the ThAr comparison lamp were obtained before and after 
every observation during the night to determine the wavelength solution.  
The spectral resolution was about 34,000 as measured by the FWHM of the ThAr emission lines in the comparison 
lamp. Exposures in each filter are summarized in Table 3. The number of objects observed changed slightly between 
observations, because fiber positions need to be reconfigured when targets pass the meridian.

Data reduction was done using standard IRAF spectroscopic packages. 
The uneven sky intensity in the CCD required special non-standard methods of sky subtraction. These procedures 
are described in \citet{meszaros01}.

\section{Stars with H$\alpha$ Emission}

\subsection{Determining H$\alpha$ Emission}

Two methods have been used for the identification of H$\alpha$ emission. 
The first selects stars with flux in the H$\alpha$ wings lying above the local continuum. 
Strong emission can be easily found, however faint emission comparable to the noise 
of the continuum can be missed. A second method was introduced by \citet{cacciari01} and is illustrated in 
Figure 2. They select a star without emission and the H$\alpha$ absorption line from this star is subtracted 
from the other spectra. With this method weak emission can be identified, but it strongly depends on the template 
selected. The H$\alpha$ absorption profile depends on temperature, as well as broadening from turbulent velocity 
and rotation, both of which could introduce features in the subtracted profile. An individual Kurucz model can 
be made for every star as a template, and the temperature problem can be avoided, but the uncertainty of other 
physical parameters can introduce similar effects in the subtracted spectrum. 
In a continuum$-$normalized spectrum,  the H$\alpha$ emission appears above the continuum level for the 
majority of stars (L465 and L72 in Figure 2). However in 
fainter stars only the shape of the H$\alpha$ line profile is changing and the emission does not appear above 
the continuum, rather just a small additional flux emerges in the absorption wings (L250 and L252 in Figure 2). 
The identification of this kind of emission can be challenging (L403 in Figure 2).
In this paper, we used both methods. However, we selected 8 stars with no emission and of different colors and 
luminosities to make the template. The stars identified with H$\alpha$ emission are the same with both methods.

No matter which method is used, the detection of faint emission depends on the reduction technique. Continuum 
normalization and sky subtraction can change the emission flux and move it above or below the continuum level 
(L252 in Figure 2). Continuum normalization was done using a low$-$order Chebyshev function in the IRAF task 
{\it continuum} in order to fit the continuum and filter throughput. The continuum placement strongly depends 
on the order of the function and the rejection limits below and above the fit.  
Sky subtraction is especially challenging with Hectochelle because additional counts appear 
between aperture numbers 100 and 150, possibly due to scattered light. This additional flux depends not only on 
the aperture but also wavelength, and although a reduction system was developed \citep{meszaros01}, all 
sky background cannot be subtracted in the middle section of the CCD. For these 
reasons the emission of very faint stars can be hard to identify and this can introduce uncertainties in 
the statistics of the presence of emission.

\subsection{H$\alpha$ Emission in the CMD}

Emission in H$\alpha$ signals an extended and high$-$temperature chromosphere; in addition the asymmetry of the 
emission indicates chromospheric mass motions \citep{dupree01, mauas01}. We observed a total of 123 different 
red giant stars in M13 and found 19 with H$\alpha$ emission. In M92, we found 9 stars with H$\alpha$ emission 
out of 64 objects. Emission above the continuum in the H$\alpha$ profile can be seen in Figures 3 and 4. For 
comparison, Figure 4 includes a star that exhibits no emission. The color-magnitude diagram (CMD) for each night 
of observation appears in Figure 5 for both clusters. The intensity ratio, B/R, of Blue (short wavelength) and 
Red (long wavelength) emission peaks for stars showing emission is contained in Table 4.

In M13, emission is found in stars brighter than V=14.69, 
corresponding to $M_V=+0.21$, using the apparent distance modulus $(m-M)_V=14.48$ from \citet{harris01}. 
The star which marks the faint luminosity limit (L719) appears to be either a blend, or a physical binary 
RGB star judging from its position in the CMD (bluer than RGB stars at the same absolute magnitude). 
This star also had a significant radial velocity change between observations (see Section 4.1). 
Among the RGB stars, L1073 at V=12.88 ($M_V=-1.60$) marks the faint luminosity limit of H$\alpha$ emission. 
Stars brighter than this are on the RGB or AGB; the evolutionary status cannot 
be determined from the CMD itself. Stars in M92 that show emission are brighter than V=14.54 ($M_V=-0.1$), 
using the apparent distance modulus $(m-M)_V=14.64$ from \citet{harris01}. However the faintest star 
(IX-12) showing emission appears to be an AGB star, according to its position in the CMD. 
Considering stars on the RGB, the star IV-94 (V=13.06, $M_V=-1.58$) appears to be the faintest RGB star 
showing emission in M92 (although the differences between the RGB and AGB at that part of the CMD are very
small). In M15, the faint luminosity limit showed significant changes between observations; this amounted to a 
change in the faint magnitude limit of 0.79 magnitudes \citep{meszaros01}. 
One can assume that the emission behaves very similarly in these clusters as well, and that the 
faint luminosity limit of H$\alpha$ emission is not constant. 

For comparison on a luminosity scale, unreddened colors for M13 and M92 stars were calculated. Foreground 
reddening [$E(B-V)=0.02$ for both clusters] and the apparent distance modulus were taken from \citet{harris01}.
The effective temperatures, bolometric corrections, and luminosities were obtained from the $V-K$ colors 
(Tables 5 and 6) using the empirical calibrations by \citet{alonso01,alonso02} and the cluster average metallicity 
[Fe/H]=$-1.54$ for M13, [Fe/H]=$-2.28$ for M92 \citep{harris01}. 
Thus on the red giant branch alone, emission appears in stars brighter than log~$(L/L_{\odot})=2.79$ in M13 and 
$\sim 78\%$ of these stars (18) show H$\alpha$ emission. In M92 this luminosity limit is slightly lower than in 
M13, log~$(L/L_{\odot})=2.74$, and also $\sim 78\%$ of these stars (7) show emission. 

Although both clusters were observed on two different days, the configurations were chosen to eliminate stars 
already observed in order to achieve full coverage of stars in the CMD. When it was possible, previously observed 
stars were configured to the remaining fibers, but the number of stars observed twice for both 
clusters is very small, only 17 in M13 and 15 in M92. Of the stars showing H$\alpha$ emission, 
comparison was possible for only two stars in M13 and three stars in M92. In M13, between 2006 March 14 
and 2006 May 10, L72 changed asymmetry (see Figure 3), while for the other star, L719, the already weak emission 
vanished. In M92, all three stars (II-53, VII-18, and IX-12) kept the same emission asymmetry, but the flux level 
of IX-12 changed in only two days (see Figure 4). 

\section{Radial Velocities}

\subsection{Cross-correlation Technique}

To measure accurate radial velocities, we chose the cross-correlation method using the IRAF task {\it xcsao}. 
Two filter regions, OB25 and RV31, were used for radial velocity measurements. The spectral region on the RV31 
filter between 5150~\AA \ and 5300~\AA \ contains several hundred narrow photospheric absorption lines of 
predominantly neutral atoms and very few terrestrial lines, thus the 
cross-correlation function is narrower than from the H$\alpha$ region, which only contains $\sim$10 lines 
(Figure 6). In the OB25 filter, the region selected for the 
cross-correlation spanned 6480~\AA \ to 6545~\AA \ purposely omitting the H$\alpha$ line.
This results in 100$-$200~m~s$^{-1}$ errors with the RV31 filter as compared to 200$-$400~m~s$^{-1}$ using the 
wavelength region earlier described in the H$\alpha$ filter. Spectra of our targets from both filters were 
cross-correlated against 2280 spectra calculated by \citet{coelho01} covering temperatures between 3500 and 
7000~K, metallicities between [Fe/H]=$-$2.5 and +0.5, and log g between 0 and 5. Radial velocities were corrected 
to the solar system barycenter. To calculate the radial velocity of a star, the radial velocities from ten 
templates with the highest amplitude of the cross-correlation function for each filter were collected and averaged 
together. A sample of the template spectra compared to an observation can be seen in Figure 6. The physical 
parameters of the templates that were used for the radial velocity measurements usually agreed with each other 
within 200~K in temperature, 1 in log~$g$, and $-$0.5 in [Fe/H] with our calculated physical parameters (Tables 5 
and 6). For almost every star the radial velocity differences among the 10 highest correlation templates in each 
filter were less than $0.5 \ \pm 0.2$~km~s$^{-1}$, which is close to the error of the individual measurements. 

We compare our results with those found in the literature. 
In M13, \citet{soderberg01} used the Hydra spectrograph on the 4$-$m Mayall telescope to obtain spectra of 150 
stars. Their template for the cross-correlation was an averaged spectrum of all giants for each Hydra observation. 
Therefore the individual radial velocities were determined as compared to the average cluster velocity. The radial 
velocity of the averaged spectrum was calculated by cross-correlating it to the solar spectrum. Comparison of the 
results can be seen in Figure 7. Errors spanned 0.5~km~s$^{-1}$ to 3$-$5~km~s$^{-1}$ in their sample, and there is 
a systematic $1.1 \ \pm 0.5$~km~s$^{-1}$ offset (Figure 7, left upper panel) between our radial velocities and 
those of \citet{soderberg01}. Hectochelle velocities determined using the H$\alpha$ region from 2006 March 14 
agreed with the observations two days later with the RV31 filter (see Figure 7, left lower panel) for the same 
stars. Radial velocities calculated from the data taken with the RV31 filter in 2006 May also agreed with data 
taken with the OB25 filter on 2006 March 14 (Figure 8, right upper and lower panel). 
We find the average radial velocity of M13 to be $-243.5 \ \pm 0.2$~km~s$^{-1}$, which is slightly 
lower than the cluster radial velocity ($-245.6 \ \pm 0.3$~km~s$^{-1}$) quoted in the \citet{harris01} catalog. 

Five stars observed with Hectochelle in M13 were reported as possible binaries by \citet{shetrone01}, when the 
radial velocities measured with the 3-m Shane telescope (Lick Observatory) were compared with velocities determined by
\citet{lupton01}. In all of these stars, differences between the two observations 
were larger than 4~km~s$^{-1}$ which exceeds the measurement errors of
$\sim$ 1~km~s$^{-1}$ and may reflect intrinsic stellar variability or binary reflex motions. Our radial
velocities differ by $4-5$~km~s$^{-1}$ compared with \citet{lupton01}, but agree within $1-2$~km~s$^{-1}$ 
with \citet{shetrone01}, which also suggests that long-term changes 
are present. Among these five stars, we observed one, L72, which showed 2.1~km~s$^{-1}$ velocity change between 
2006 March and 2006 May. \citet{lupton01} identified 
this star in M13 as a possible binary from variations in radial velocity over several years of observations. 
L72 is also known as a pulsating variable star with a possible period of 41.25 days \citep{russeva01}, so the 
velocity change found here may also relate to pulsation. L719, which marks the faint luminosity limit of stars 
showing H$\alpha$ emission, also had radial velocity changes 
\footnote{If this object were a single line binary, the
  velocity change allows only lower limits to the period (P$>$90 days)
  and semi-major axis (a $\ga$10R$_\star$). The putative companion to
  the red giant could be either a white dwarf or a late main sequence
  star, and probably the former since the color is bluer than a red giant.} 
between 2006 March 14 to 2006 May 10 from $-254.1\ \pm 0.3$~km~s$^{-1}$ to $-245.2\ \pm 0.2$~km~s$^{-1}$.
No other stars showed significant, larger than 2~km~s$^{-1}$, variations in our sample in M13.

A large sample of stars in M92 was observed by \citet{drukier01} using the HYDRA
multi-fiber spectrograph on the 3.5-m WIYN telescope. Their errors spanned 0.3$-$1.2~km~s$^{-1}$.
The comparison of results can be seen in Figure 8. Radial velocities for the same stars agreed within
the errors (Figure 8, left upper panel). The Hectochelle spectra give
the cluster average radial velocity as $-118.0 \ \pm 0.2$~km~s$^{-1}$, which is lower than the value  
($-120.3 \ \pm 0.1$~km~s$^{-1}$) quoted in the \citet{harris01} catalog. In M92, two stars show radial 
velocity variations, which usually indicates binarity or pulsation. 
II-53 had a significant velocity variation of 7.7~km~s$^{-1}$ between 2006 May 7 
and 2006 May 9. Another star, XI-38, showed a 4.9~km~s$^{-1}$ difference between the radial
velocity measured by \citet{drukier01} and the velocity measured by us
on 2006 May 7.

\subsection{Bisector of H$\alpha$ Lines}

To search for mass motions in the chromosphere, we evaluated the H$\alpha$ core asymmetry using a bisector method. 
The difference between the center of the line core and 
the line center near the continuum level gives a measure of the atmospheric dynamics through the chromosphere. 
To accomplish this, the line profile with continuum normalization was divided into 20 sectors. The top sector was 
usually between 0.7 and 1.0 of the continuum in the normalized spectrum, the lowest sector was placed 
0.01~$-$~0.05 above the deepest part of the line depending on its signal-to-noise ratio. 
The velocities of the H$\alpha$ bisector asymmetry ($v_{bis}$) are calculated in the following way: the top and 
the bottom 3 sectors are selected, the wavelength average of each sector is calculated, then subtracted one from 
another and changed to a velocity scale. The bisector velocities, $v_{bis}$, are shown in Figure 9 and listed in 
Tables 7 and 8. A negative value corresponds to an outflowing velocity. The error of
the majority of the measurements spanned 
0.5$-$1.0~km~s$^{-1}$; only stars fainter than V=15 magnitude exceeded 1.0~km~s$^{-1}$ in measurement error. 
RGB stars fainter than log~$(L/L_{\odot})=2.5$ did not show asymmetry in the H$\alpha$ core and $v_{bis}$ was 
nearly zero. This luminosity is nearly the same for both clusters and also very similar to M15 \citep{meszaros01}, 
which suggests that the luminosity limit of the line core asymmetry marking the onset
of expansion does not depend on average cluster metallicity. 
Stars brighter than log~$(L/L_{\odot})=2.5$ showed core asymmetry and the majority of the bisectors were blue 
shifted. The start of chromospheric outflow presumably relates to mass loss. However, the value of 
$v_{bis}$ appears to depend on luminosity. In the metal-rich cluster M13, $v_{bis}$ increases with
luminosity and reaches its maximum value ($\approx $~5~km~s$^{-1}$) at about log~$(L/L_{\odot})=3.2$ but the most 
luminous stars exhibit lower (near zero km~s$^{-1}$) values. 
In the more metal-poor cluster M92, $v_{bis}$ also increases with
luminosity and reaches the maximum value at log~$(L/L_{\odot})=3.4$ but the decrease in outflow velocity is much 
smaller. At the same luminosity, T$_{eff}$ for the metal-rich M13 giants is lower than for M92. Thus, the apparent 
decrease in $v_{bis}$ at high luminosity would appear to be related to the changing atmospheric structure 
(see Section 6.1).

Where AGB stars are well separated in color from 
the RGB stars in M92, the AGB stars exhibit larger values of $v_{bis}$ 
than RGB stars (Figure 10). The star 
IX-12 in M92 shows the largest bisector velocity at $v_{bis}=$$-15.9\ \pm 1.3$~km~s$^{-1}$, 
and this star is the faintest star showing emission in the cluster (see Figure 9). 
Its position in the CMD suggests that this star is an AGB star (Figure
5). AGB stars in M13 generally have lower bisector velocities than
AGB stars in M92 which suggests that the metal-rich objects have
slower winds.

\section{\ion{Ca}{2} K Profiles}

Spectra in the \ion{Ca}{2}~H$\&$K region were obtained for 119 red giant stars in M13 and 63 red giants in M92. 
The profiles of the \ion{Ca}{2}~K core ($\lambda$3933) are shown in Figures 11 and 12 for all stars exhibiting emission. 
The position of these stars in the CMD can be seen in Figure 13. The
intensity ratio of the emission core,\footnote{B signifies the short-wavelength emission peak
  and R the long-wavelength emission peak.}
B/R, is summarized in Table 9. Because of the high radial velocity of M13 ($v_{rad}$=$-$243.5~km~s$^{-1}$), 
absorption by the local interstellar medium (ISM) is well away from the \ion{Ca}{2} K core. 
Although the ISM is closer to the \ion{Ca}{2} K emission in M92 ($v_{rad}$=$-$118.0~km~s$^{-1}$), 
it does not affect the emission profile. 
Red giant stars have low flux levels near 3950~\AA, hence the deep photospheric absorption in 
\ion{Ca}{2}~H$\&$K causes the photon noise to become comparable to the flux of the core emission for stars 
fainter than 14th magnitude. Determining the presence of emission is challenging. 
The spectra of our targets were compared to a Kurucz model, [computed by \citet{coelho01} without a chromosphere], 
to verify the emission. Altogether 34 stars showed \ion{Ca}{2}~K emission in M13 and 12 in M92. 

The spectra of H$\alpha$ and \ion{Ca}{2} were obtained on the same night or separated by 1 or 2 days, and the 
asymmetry of the K-line emissions is similar to that found in H$\alpha$ for the majority of the stars. 
The brightest stars in M13 showed stronger \ion{Ca}{2}~K emission than stars in M92 at the same luminosity. This results 
because the lower effective temperatures of M13 giants increase the contrast of the emission to the continuum.  
For stars fainter than V=14, the ratio of the blue to the red side of the 
\ion{Ca}{2}~K emission core, or even the presence of the emission is difficult to determine.

The velocity of the central reversal (K$_3$) was measured for the brightest stars (Table 10) using three strong 
absorption lines closest to \ion{Ca}{2}~K as a photospheric reference. A Gaussian function using the IRAF task 
{\it splot} was fitted to the cores of the photospheric lines and the central reversal absorption of 
\ion{Ca}{2}~K. Radial velocities of the photospheric lines were averaged and then subtracted from the radial
velocity of the \ion{Ca}{2}~K$_3$ feature. The velocity shift of the K$_3$ absorption lies between 0~km~s$^{-1}$ 
and $-$16~km~s$^{-1}$ (Table 10). 

\section{Discussion}

In this paper and in our previous study \citep{meszaros01}, we have presented H$\alpha$ and 
\ion{Ca}{2}~K spectroscopy of 297 red giants in 3 globular clusters with different metallicities 
\citep{harris01}: M13 ([Fe/H]=$-$1.54), M15 ([Fe/H]=$-$2.26), and M92 ([Fe/H]=$-$2.28). The presence of 
emission in these transitions signals an extended, high-temperature chromosphere, and the asymmetry of the 
emission and the line core indicates chromospheric mass motions. Comparison of the statistics of the  profile 
characteristics among the globular cluster stars could reveal the effects of metallicity on mass loss. In the 
following sections, we discuss several  characteristics of the presence of
H$\alpha$ and \ion{Ca}{2} emission and the resulting velocity signatures. Parameters of the clusters and 
the lines can be found in Table 11. In the final section of the Discussion, we compare our
results with those of \citet{cacciari01} who presented similar line profiles for 137 red giants in 
the globular cluster NGC~2808.

\subsection{Presence of H$\alpha$ Emission}

On the RGB, H$\alpha$ emission sets in  for all stars with $T_{eff} \la 4500$ K and 
log~$(L/L_{\odot})\ga 2.75$ in all 3 clusters: M13, M15,\footnote{In M15, the stars K672 and K875 are 
clearly AGB objects, and the faintest RGB star showing emission in that study is B30.}  and M92. 
It is perhaps fortuitous that the limits are so similar since the presence of H$\alpha$
can change by as much as 0.79 magnitudes from observations on one date to another \citep{meszaros01}. 
Stars on the AGB show H$\alpha$ emission to lower luminosity limits than the RGB objects. The faint limits of 
emission for AGB stars in M13 and M92 are comparable, while AGB stars in M15 with emission are
brighter. Emission is variable in all giants and again there does not appear to be a systematic dependence of 
luminosity limits on metallicity. This result suggests that whatever mechanism produces the variable emission
occurs similarly in all metal-poor red giants.

The percentage of stars showing inflow and outflow in the H$\alpha$ emission wings\footnote{As measured by 
the ratio of B/R -- the short wavelength to long wavelength 
peaks of the H$\alpha$ emission wings.} varies from cluster to cluster and appears to be related to cluster 
metallicity. In the metal-poor M92, about 82$\%$ of stellar spectra showing
emission have an inflow signature (18$\%$ show outflow) and the study of M15 
\citep{meszaros01} revealed that about 78$\%$ of stars with H$\alpha$ emission displayed an inflow signature 
(22$\%$ outflow). These two clusters show similar behavior in their chromospheric dynamics.  
M13 has a more equal distribution of the dynamical signature with 55$\%$ of the H$\alpha$
spectra indicating inflow (45$\%$ outflow). Since all the luminous stars are losing mass,
it might be puzzling why the dominant emission signature in H$\alpha$ is one of inflow.
(And as discussed in the following section, the line cores generally
indicate outflow.) We believe this relates to the dynamics of the atmosphere.
It appears likely that our targets are pulsating \citep{mayor01}, and comparison 
with the well-known pulsators, Cepheids, shows that inflow signatures
in H$\alpha$ are accompanied by inflow in the metallic photospheric lines \citep{baldry97},
and that inflow occurs for a larger fraction of the pulsation period in longer period Cepheids than in 
shorter period stars \citep{nardetto01, petterson01}. In this way, we understand the
dominance of inflow emission signatures. The different proportions of inflow/outflow signatures 
may indicate that the pulsation period in M15 and M92 is generally longer than in M13. It is clear that
variability is ubiquitous. Almost all stars brighter than V=12.5 are variables in M13, but only one variable 
red giant exhibits periodic photometric variations, and the remaining 
ones are semi-regular or irregular \citep{kopacki01}. 

The fraction of stars showing H$\alpha$ emission increases with luminosity and decreasing effective temperature. 
Because of the separation of red giant branches in the CMD due to metallicity, the distribution of the emission 
with luminosity and effective temperatures differs between clusters. At the same luminosity, the metal rich M13 
exhibits more stars with H$\alpha$ emission than the metal poor M92 or M15, because the effective
temperatures are lower in M13 for a constant luminosity. However, at the same stellar effective temperature,
M15 and M92 exhibit more stars showing emission than M13. This may appear as a metallicity effect, but it 
originates in the fact that both high luminosity and low effective temperature produce more H$\alpha$ 
emission.

\subsection{The H$\alpha$ Bisector Velocity}

Stars brighter than log~$(L/L_{\odot})=2.5$ show a blue-shifted H$\alpha$ core in both M13 and M92, and 
outflows become faster with increasing luminosity (Figure 9). This occurs in M15 as well \citep{meszaros01}. 
Thus, the luminosity at the onset of outflow, indicated by the H$\alpha$ line core is independent of metallicity. 
The behavior of the bisector velocity on the RGB changes at the highest luminosities 
(Figure 14). Giants in M13, the most metal-rich cluster, show lower bisector velocities
in the most luminous (and coolest stars). In fact, the velocities of the H$\alpha$ core approach 0~km~s$^{-1}$ 
with respect to the photosphere, signaling that the outward motions have decreased in the atmospheric region where 
the H$\alpha$ line forms. Since the brightest stars in M13 have a lower T$_{eff}$ than those in M15 and M92 
(Fig. 14), we suspect that the decrease in the H$\alpha$ bisector velocity results from 
the changing structure of the very extended atmosphere. The H$\alpha$ wing asymmetry and \ion{Ca}{2}~K asymmetry
preponderantly signal outflow in the most luminous stars. It is noteworthy that outflow begins at a luminosity, 
log~$(L/L_{\odot})\sim 2.5$, and as the stars become more luminous, emission wings occur in
the H$\alpha$ profile in our sample at log~$(L/L_{\odot})\sim 2.75$ . We conjecture that the onset of pulsation
marked by the observed outward motion leads to a warmer chromosphere producing emission
wings in H-$\alpha$.

Differences in the coreshift between AGB and RGB stars can be seen where these stars are distinct on the CMD. 
Stars on the AGB, between log~$(L/L_{\odot})=2.0$ and 2.7, showed slightly larger bisector velocities 
than RGB objects in both clusters, although the values are most extreme in M92. 
AGB stars tend to have lower values of log g and smaller escape velocities in the chromosphere as compared 
to RGB stars, which makes them more sensitive to mass loss driving mechanisms resulting in faster winds. 
We can speculate that there is more heating in the hotter AGB stars; it may be that a putative magnetic 
field is stronger after the stars have been through helium burning enabling enhanced wave motions, 
heating, and acceleration in the chromosphere as compared with RGB stars. The extremes of outflow velocity on the 
AGB tend to be larger in the metal-poor clusters, M15 and M92, than in M13 (see Figure 10 and 14). 

There is no evidence here that the outflow velocity is slower at low metallicity. This suggestion resulted from 
observations of OH/IR stars in the low metallicity Magellanic Clouds
\citep{marshall04}, and modeling of dust driven winds \citep{helling01, wachter01} such as those 
identified in Omega Centauri by Spitzer Space Telescope observations \citep{boyer01, mcdonald02}. In fact,
we find just the opposite. M15 and M92 have generally higher 
velocities than stars in M13 (see Figure 14, lower panels). No evidence for a `super-wind' \citep{renzini01, bowen01} in 
the sense of an abrupt high velocity outflow is present in our spectra, although the largest mass-loss 
rates might be expected for stars with lower T$_{eff}$ than found in this sample \citep{wachter02}.
Even in the dusty red giants in M15 detected with the Spitzer Space Telescope \citep{boyer02}, the bisector 
velocities have similar values as red giants without an IR excess \citep{meszaros01}. 
These similarities suggest that mass loss and dust production are not
correlated, and the triggering of dust production may be an episodic phenomenon. 

Three stars in M13 exhibit large ($>$2~km~s$^{-1}$) bisector-velocity changes between observations. 
The star at the lowest luminosity limit, L719, shows a 4.7~km~s$^{-1}$ bisector velocity difference,
which is clearly visible on the spectrum (see Fig. 3) as the H$\alpha$ emission disappeared.  
K342 and K658 showed 2.2~km~s$^{-1}$ and 4.9~km~s$^{-1}$ changes respectively, but these stars are faint and 
the error due to the low signal-to-noise of the spectra is comparable to the velocity change.  
In M92, only 2 stars show a large coreshift in H$\alpha$: II-6 is a very faint star and this difference is 
comparable to the error of our measurements; IX-12 is an AGB star and shows similar $v_{bis}$ values 
as stars of the same luminosity. However between our 2
observations separated by 2 days, the coreshift changes for other stars are relatively small. In M15 
\citep{meszaros01}, it is the AGB stars in the log~$(L/L_{\odot})=2.3-3.0$ luminosity region that show large 
bisector velocity changes (3$-$7~km~s$^{-1}$) over a time span of one year or more. 

On the RGB, the velocities in M15 indicate that systematic outflow (more negative than $-$2~km~s$^{-1}$) in the 
H$\alpha$ core occurs at  luminosities, log~$(L/L_{\odot}) \ga 2.5$. 
The velocity of the outflow increases with luminosity and only the brightest stars show slightly smaller outflow 
(Figure 9 and 14). In M92, which is also very metal-poor, stars brighter than 
log~$(L/L_{\odot})=2.5$ showed bisector velocities more negative than $-$2~km~s$^{-1}$, and only the brightest 
star shows smaller outflow velocities. If there are differences in mass loss between M15 and M92, the shapes of 
the H$\alpha$ line profiles do not reflect this. Thus the H$\alpha$ line by itself cannot help
to decide if mass loss is the `second parameter' in M15 and M92. \citet{mcdonald01} found no significant 
correlation between core asymmetry and luminosity, when they examined H$\alpha$ and \ion{Ca}{2} IR triplet 
spectra of 47 red giant stars near the RGB tip in 6 globular clusters. Above a 
certain luminosity the bisector velocity of H$\alpha$ becomes small and motions are difficult to detect in this 
region of the atmosphere, independent of cluster metallicity. Possibly another diagnostic such as 
\ion{He}{1} $\lambda$ 10830 or ultraviolet lines, formed
higher in the atmosphere needs to be examined. A \ion{He}{1} $\lambda$10830 absorption 
line was detected \citep{smith02} in the star L687 in M13, with an extension to $-$30~km~s$^{-1}$, suggesting 
that when the helium line becomes detectable (apparently for $T_{eff} \sim$ 4650~K), it can give an 
indication of a faster wind. This \ion{He}{1} $\lambda$10830 velocity is comparable 
to the higher values of the H$\alpha$ bisector velocities that are found here in AGB stars. 
\citet{pilachowski01} classifies L687 as an AGB star.

\subsection{The \ion{Ca}{2} Emission and Velocity}

The \ion{Ca}{2}~K$_{2}$ emission appears at lower luminosities than H$\alpha$ emission in both clusters 
[log~$(L/L_{\odot})=1.92$ for M13 and log~$(L/L_{\odot})=1.96$ for M92]. In M15 
\citep{meszaros01} the \ion{Ca}{2} K luminosity limit agrees with the H$\alpha$ emission limit 
[log~$(L/L_{\odot})=2.36$], but the low signal-to-noise ratio of those observations did not
allow us to determine the presence of \ion{Ca}{2}~H$\&$K emission in fainter stars. The lower luminosity limit 
of \ion{Ca}{2}~K emission does not appear to be dependent on the cluster metallicity. 

The number of stars with \ion{Ca}{2} emission in both M13 and M92 exceeds the number of stars showing H$\alpha$ 
emission. Stars with H$\alpha$ emission generally have \ion{Ca}{2}~K emission,
but not all stars with \ion{Ca}{2}~K emission show H$\alpha$ emission. This difference is not unexpected 
since the regions of formation of the \ion{Ca}{2}~K core and the H$\alpha$ emission wings are separated in the 
atmosphere of a giant. Models suggest that \ion{Ca}{2}~K emission forms lower in the atmosphere than H$\alpha$ 
wings \citep{dupree03}. Additionally H$\alpha$ shows variations in asymmetries 
over the span of a few days \citep{cacciari02, meszaros01} which could contribute 
to the differences. Some stars in both clusters were observed twice at 
\ion{Ca}{2}. Changes in \ion{Ca}{2}~K emission were observed in  
two stars in M13 and four stars in M92 where the line profile of \ion{Ca}{2}~K, either changed asymmetry or 
the emission strengthened or weakened, or both. 

The outflow velocities of the \ion{Ca}{2} K$_3$ reversal are generally larger than the bisector velocities 
of the H$\alpha$ line for the same stars (Figure 9). Similar behavior was found by \citet{zarro01} in 
Population I giants and supergiants, and they concluded from the similarity of
line profiles that \ion{Ca}{2}~K line is formed higher in the atmosphere and the increased outward velocity 
reflects a mass-conserving outflow. While models of the Sun suggest that the \ion{Ca}{2}~K$_3$ feature forms in a 
higher atmospheric region than the core of the H$\alpha$ line \citep{avrett01}, some chromospheric models of
metal-deficient giants \citep{dupree03, mauas01} locate the approximate depth of formation of the 
\ion{Ca}{2}~K$_3$ feature below that of the H$\alpha$ core. These models would suggest the opposite conclusion 
from Population I stars, that the flow is decelerating in the upper atmosphere.
Yet another model \citep{dupree04} for the metal-deficient giant HD 6833 finds the contribution function of 
\ion{Ca}{2}~K$_3$ to lie above that of the H$\alpha$ core and hence signal an accelerating outflow. 
Some ambiguity may exist in the definition of the region of formation, and in addition it can extend over a 
substantial height in the chromosphere. In some cases, the \ion{He}{1} $\lambda$ 10830 line, clearly formed
above both \ion{Ca}{2}~K and H$\alpha$ shows even higher outflow velocities in metal-deficient stars 
\citep{dupree04, smith02}, so the accelerating outflow models appear preferable.

\subsection{Comparison to NGC 2808}

Red giants in another metal rich cluster, NGC~2808, were studied by \citet{cacciari01} and can be compared  
to M13. However, this comparison may be somewhat compromised since NGC~2808 has an extreme case of peculiar 
horizontal branch morphology \citep{lee05} and a split main sequence with potentially 3 populations 
\citep{piotto07}, making it  one of the most persuasive clusters for the existence of
possible multiple stellar populations including a super helium-rich component \citep{lee05,dantona08}. These 
features of NGC~2808 make it quite different from  M13 -- the metal rich cluster in our
sample. M13 has a more chemically homogeneous population, although it is conjectured to consist of 
predominantly second generation stars \citep{dantona08}. The average metallicity of
NGC~2808 ([Fe/H]=$-$1.15) is higher than M13 ([Fe/H]=$-$1.54), by a factor of 2.5.

The detection threshold for H$\alpha$ emission on the RGB in NGC~2808, of
log~$(L/L_{\odot})=2.5$ is fainter by $\sim$0.2 magnitudes than the limits for M13, M15,  and M92 (see Table
11). Since the appearance of emission in the H$\alpha$ line varies with
time, these limits are comparable, one with another. However, the percentage of red giants exhibiting emission 
is less at 52$\%$ than we find for M13, M15, and M92 where the value is about 80$\%$. 
The atmospheres of the NGC~2808 giants may be at lower temperatures since, for the same input energy, radiation
losses are greater due to the increased abundance of metals than in  metal-poor objects.

Differences arise in the H$\alpha$ outflow signature indicated by the emission
wings in NGC~2808 where  an exceptionally low percentage (at 7$\%$) is found by \citet{cacciari01} in contrast 
to the 45$\%$ of red giants showing outflow in M13 and 18$-$22$\%$ in the metal-poor clusters, M15 and M92.

\citet{cacciari01} measured significant core shifts ($<-$2~km~s$^{-1}$) of H$\alpha$ in 7 stars of their sample of 
giants in NGC~2808. These stars are brighter than log~$(L/L_{\odot})=2.9$.  It is interesting that the 3 most 
luminous stars in their sample had core shifts of 1~km~s$^{-1}$ or less, similar to our results for M13 (Figure 9). 
M4, another cluster of similar metallicity as NGC~2808 also did not have coreshifts (more negative 
than $-$2~km~s$^{-1}$) either in H$\alpha$ or Na~D in any of $\approx$10 stars that have luminosities 
log~$(L/L_{\odot})>3.3$ \citep{kemp01}.

The luminosity limit in NGC~2808 \citep{cacciari01} for \ion{Ca}{2}~K emission lines is log~$(L/L_{\odot})\sim2.6$ 
which is higher than  the H$\alpha$ limit in NGC~2808. This result is puzzling since \ion{Ca}{2}~K is found at   
lower luminosities than H$\alpha$ emission in the other clusters, M13, M15, and M92. The resolution of the 
Calcium spectra studied by \citet{cacciari01} was the lowest of all their spectra at R=19600, and the 
signal-to-noise in the line core for  the brightest stars was only 15. So it is 
possible  that Calcium was not detected in the fainter targets. The limit for \ion{Ca}{2}~K in NGC~2808 is 
$\sim$0.2 magnitudes brighter than found in M15 which is a metal-poor cluster. The 2 metal poor
clusters, M15 and M92 differ in the \ion{Ca}{2}~K  limit by 0.4 magnitudes. At present, there is not sufficient 
evidence to claim that the B/R ratio of \ion{Ca}{2}~K emission varies night to night as the B/R ratio of 
H$\alpha$ emission. The core shift of \ion{Ca}{2} K in NGC~2808 is generally more negative than the value for 
H$\alpha$, similar to that  found here for the most luminous stars (Figure 9). 

Since both the H$\alpha$ emission and emission wing asymmetries are variable, it is difficult to draw 
firm conclusions about systematic differences between clusters. In the sample of red giants in NGC~2808 
studied by \citet{cacciari01}, a lower fraction of stars was found with H$\alpha$ emission  and with 
outflow signatures in the emission wings than in the more metal-poor clusters (M13, M15, M92). However, 
the dynamical characteristics including the luminosity onset of outflow and wind speeds, appear 
indistinguishable among these clusters.

\section{Conclusions}

Summarizing, we find the following:

1. Hectochelle spectra of M13, M15, and M92 show H$\alpha$ emission to 
occur on the red giant branch in stars with T$_{eff} \la$ 4500 K and 
log(L/L$_\sun$)~$\ga$~2.75. AGB stars exhibit H$\alpha$ emission to 
lower luminosities.   \ion{Ca}{2} K emission extends to lower
luminosities than H$\alpha$ both on the RGB and AGB.

2. Considering 3 clusters, spanning 
[Fe/H]=$-1.54$ (M13), to [Fe/H]=$-$2.3 (M15, M92), we find no {\it systematic} 
dependence of the presence of H$\alpha$ or \ion{Ca}{2}~K emission from red giants  on
T$_{eff}$, L/L$_\sun$, or cluster metallicity.  

3. Asymmetric H$\alpha$ cores show that chromospheric material is flowing out from 
stars brighter than log~$(L/L_{\odot})\sim2.5$ and 
the speed of the outflow increases with increasing stellar luminosity.
This outflow may represent the onset of mass loss, and the luminosity
at which outflow begins  is similar for all metallicities. H$\alpha$
velocities on the red giant branch are similar for all 
metallicities (but not for AGB stars, see below). 

4. Stars on the asymptotic giant branch 
near log~$(L/L_{\odot})\sim2.0-2.7$ show higher outflow velocities than RGB stars, 
and faster outflows 
are found in the metal$-$poor M15 and M92 than the metal$-$rich M13 objects.   

5. The sensitivity of H$\alpha$ to mass motions 
decreases for $T_{eff} <$ 4000~K causing the coolest giants
in M13 to exhibit little or no outflow in this line. 

6. The \ion{Ca}{2} K$_{3}$ absorption features exhibit higher velocities than
H$\alpha$ suggesting accelerating outflows in the chromospheres.

7. We find no differences in chromospheric signatures in the profiles
   or the presence  of  H$\alpha$ and \ion{Ca}{2} that can resolve the 'second-parameter'
   problem for the paired clusters, M15 and M92. 

\acknowledgments{We thank the referee for very detailed comments that significantly improved the presentation.  
Observations reported here were obtained at the MMT Observatory, a joint 
facility of the Smithsonian Institution and the University of Arizona. 
We are grateful to the scientists at CfA who are developing and characterizing Hectochelle: Nelson Caldwell,
Daniel G. Fabricant, G{\'a}bor F{\^u}r{\'e}sz, David W. Latham, and Andrew Szentgyorgyi. The authors also would like to thank John
Roll and Maureen A. Conroy for developing SPICE software, and Mike Alegria, John McAfee, Ale Milone, Michael Calkins and 
Perry Berlind for their help during the
observations. Kyle Cudworth kindly provided coordinates and photometry for M13 and M92 stars. 
Szabolcs~M{\'e}sz{\'a}ros is supported in part by a SAO Predoctoral Fellowship, NASA, and the Hungarian OTKA 
Grant TS049872, T042509 and K76816. 
This research is also supported in part by the Smithsonian Astrophysical Observatory.}

\clearpage

\begin{figure}
\includegraphics[width=4.5in,angle=270]{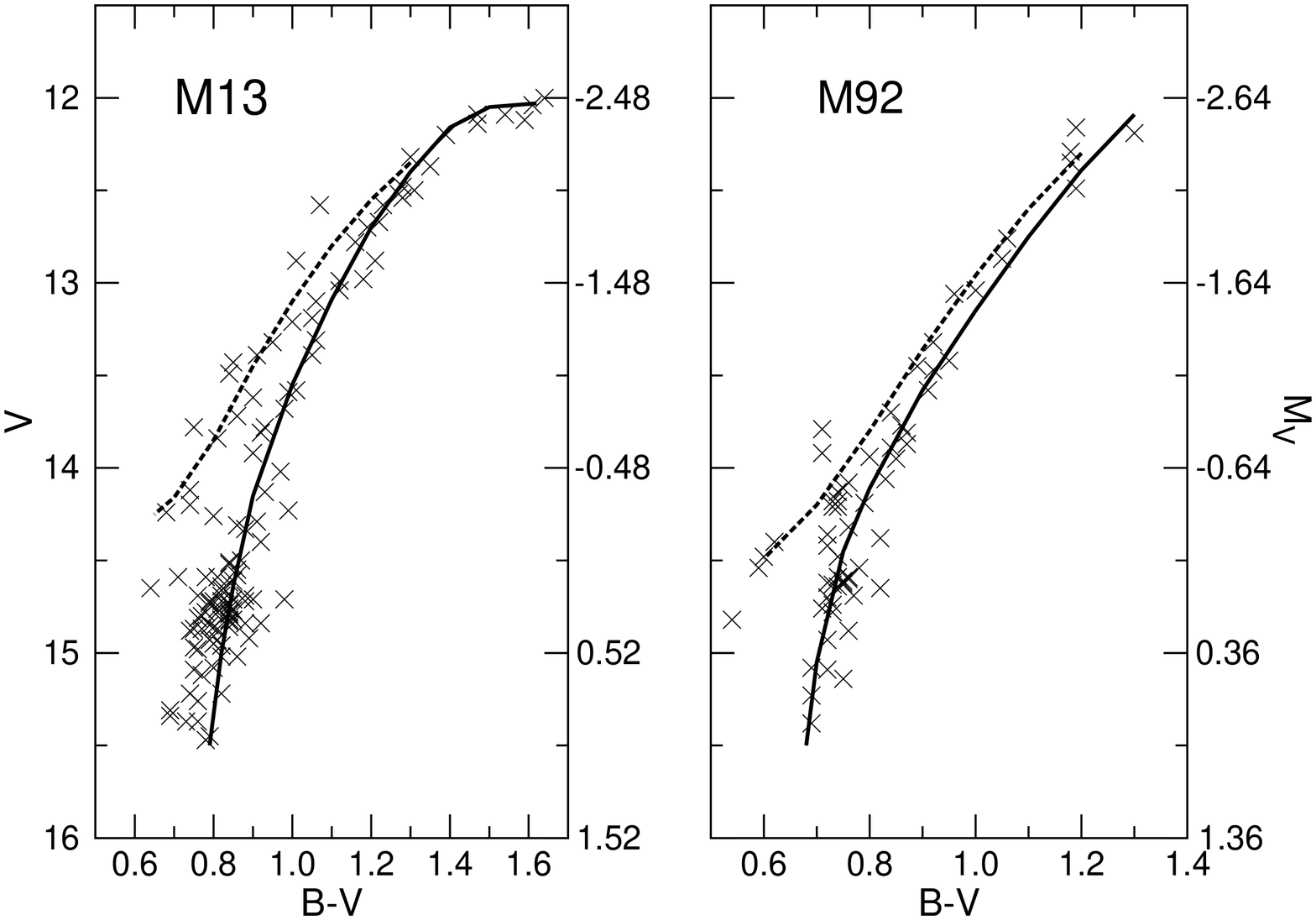}
\caption{Color-magnitude diagram for all stars observed in M13 and M92. 
The solid line shows the fiducial curve of the RGB; the dashed line shows the fiducial curve of the 
AGB for both clusters taken from observations of \citet{sandage01}. The absolute magnitudes were calculated using the 
apparent distance modulus $(m-M)_V=14.48$ for M13 and $(m-M)_V=14.64$ for M92 from \citet{harris01}.
}
\end{figure}

\clearpage

\begin{figure}
\includegraphics[width=4.5in,angle=270]{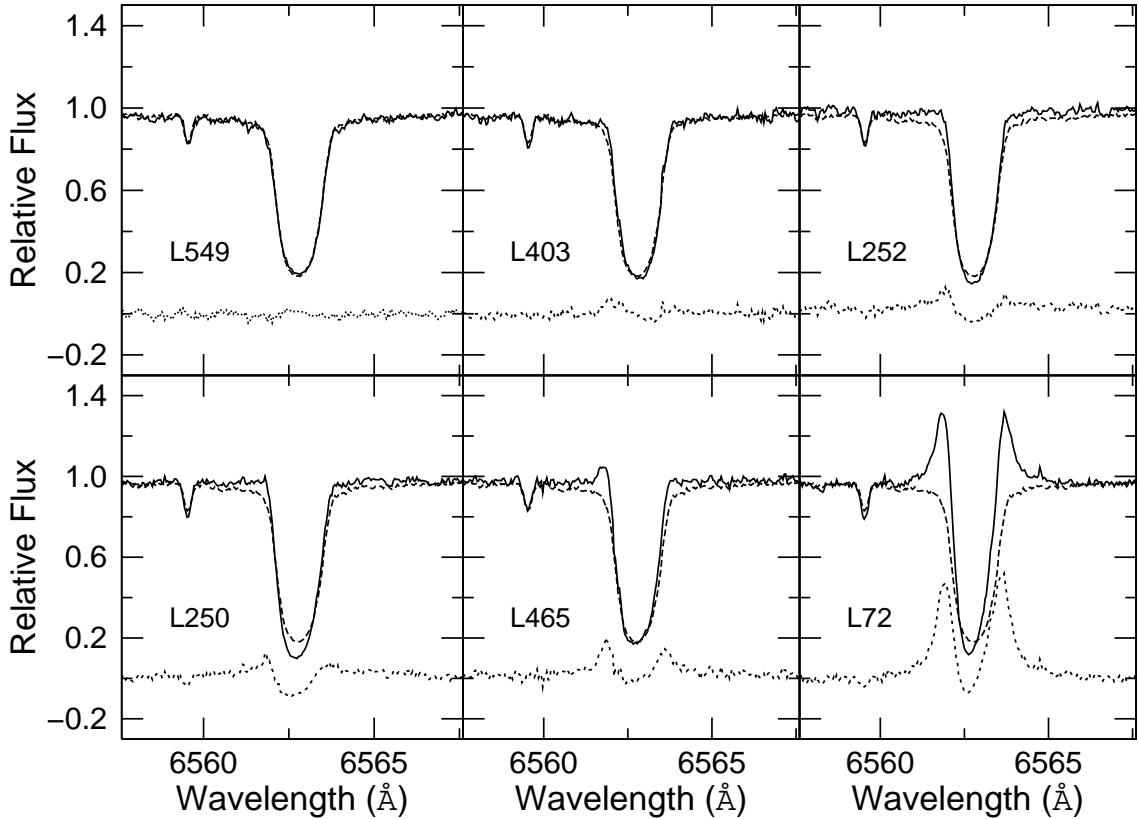}
\caption{Examples of the effect of subtracting an averaged spectrum (dashed line) from the observed spectra (solid lines)
for stars in M13. The difference spectrum is shown by a dotted line below. 
L549 is a star without any emission; the error of the subtracted spectrum (dotted line) is smaller than 0.02 of the
continuum level. In the case of
L403, weak emission on the short wavelength is visible, however it is comparable to the noise of the observed spectrum and
extends to the core of the line, 
so it was not identified as emission. L252 is an example of how the continuum normalization can shift the region near 
H$\alpha$ making it hard to identify the emission. The blue emission in the spectrum of L250 might be missed by eye, but 
the subtraction method clearly shows the presence of emission. L465 and L72 are examples of emission that is clearly 
visible in the spectrum above the continuum level.}
\end{figure}

\clearpage

\begin{figure}
\includegraphics[width=5in,angle=0]{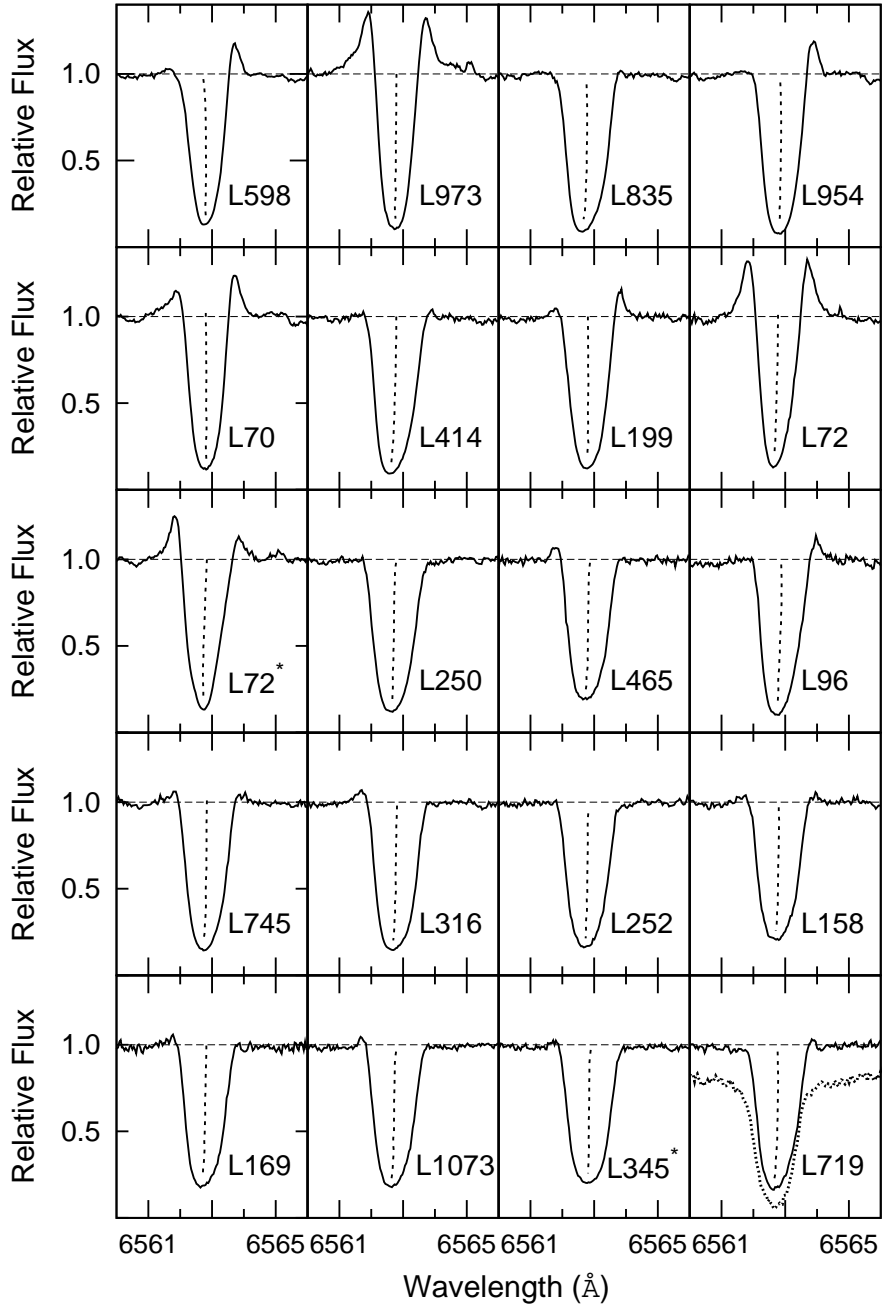}
\caption{Normalized spectra of red giants in M13 which showed emission in H$\alpha$ on 2006 March 14. 
Stars marked with $*$ were observed on 2006 May 10. The dashed line 
marks the bisector. The emission of one star, L719, disappeared
between observations, and the spectrum is 
overlaid here using  a dotted
line. The spectra are arranged in order of decreasing brightness; the brightest is at the top left 
and the stars become fainter from left to right. The wavelength scale is corrected for heliocentric velocity. 
The radial velocity of M13 is $-243.5 \ \pm 0.2$~km~s$^{-1}$.
}
\end{figure}

\clearpage

\begin{figure}
\includegraphics[width=5in,angle=0]{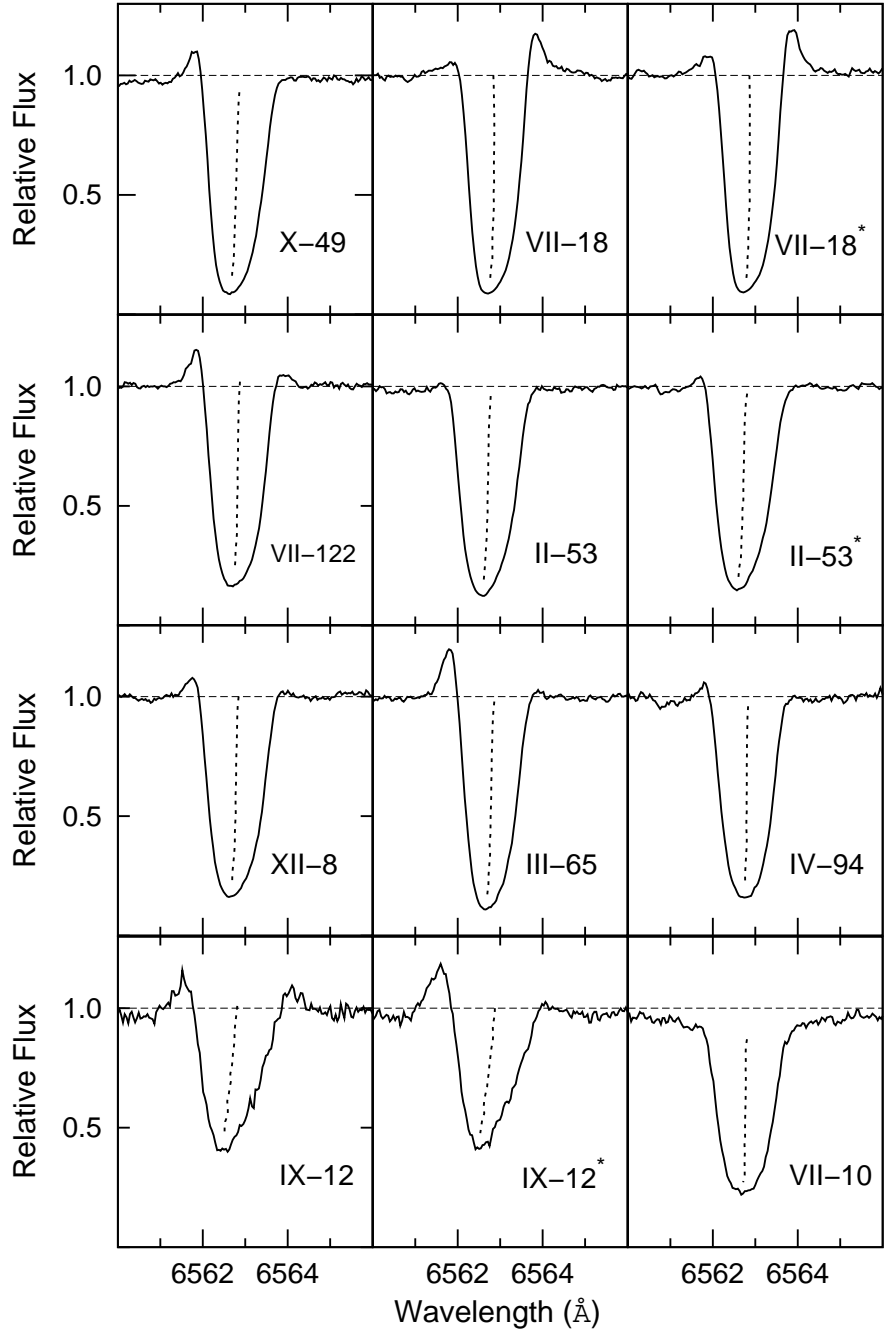}
\caption{Normalized spectra of red giants in M92 which showed emission in H$\alpha$ on 2006 May 7. 
Stars marked with $*$ were observed on 2006 May 9. VII-10 is an example of an H$\alpha$ profile without emission. 
For explanation please see Figure 3 caption. The radial velocity of M92 is $-118.0 \ \pm 0.2$~km~s$^{-1}$.}
\end{figure}

\clearpage

\begin{figure}
\includegraphics[width=4.5in,angle=270]{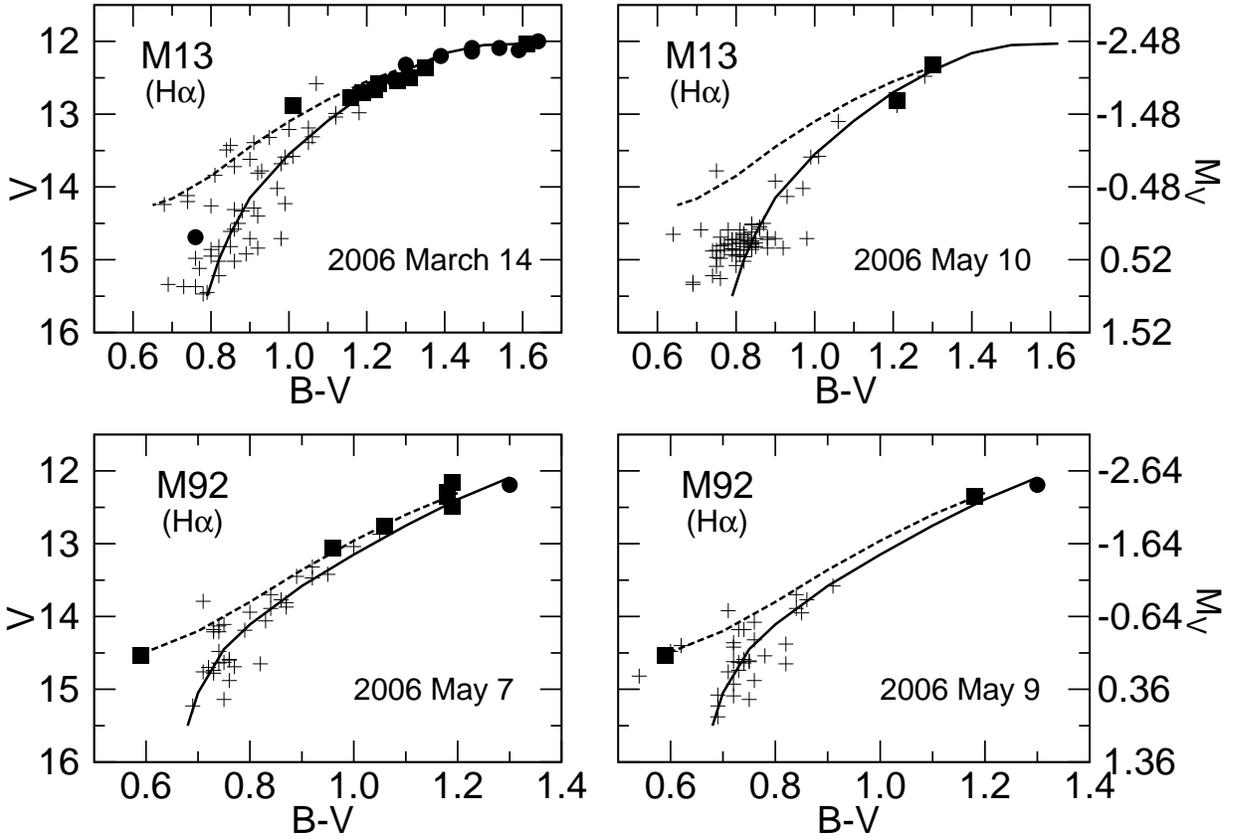}
\caption{Color-magnitude diagrams for all M13 and M92 stars observed in 2006. 
Stars with H$\alpha$ emission and with B$<$R (indicating outflow) are marked with filled circles; 
stars with B$>$R emission wings (suggests inflow) are denoted by filled squares. 
The solid line shows the fiducial curve of the RGB; dashed lines show the fiducial curve of the 
AGB for M13 and M92 from observations of \citet{sandage01}.}
\end{figure}

\clearpage

\begin{figure}
\includegraphics[width=5in,angle=0]{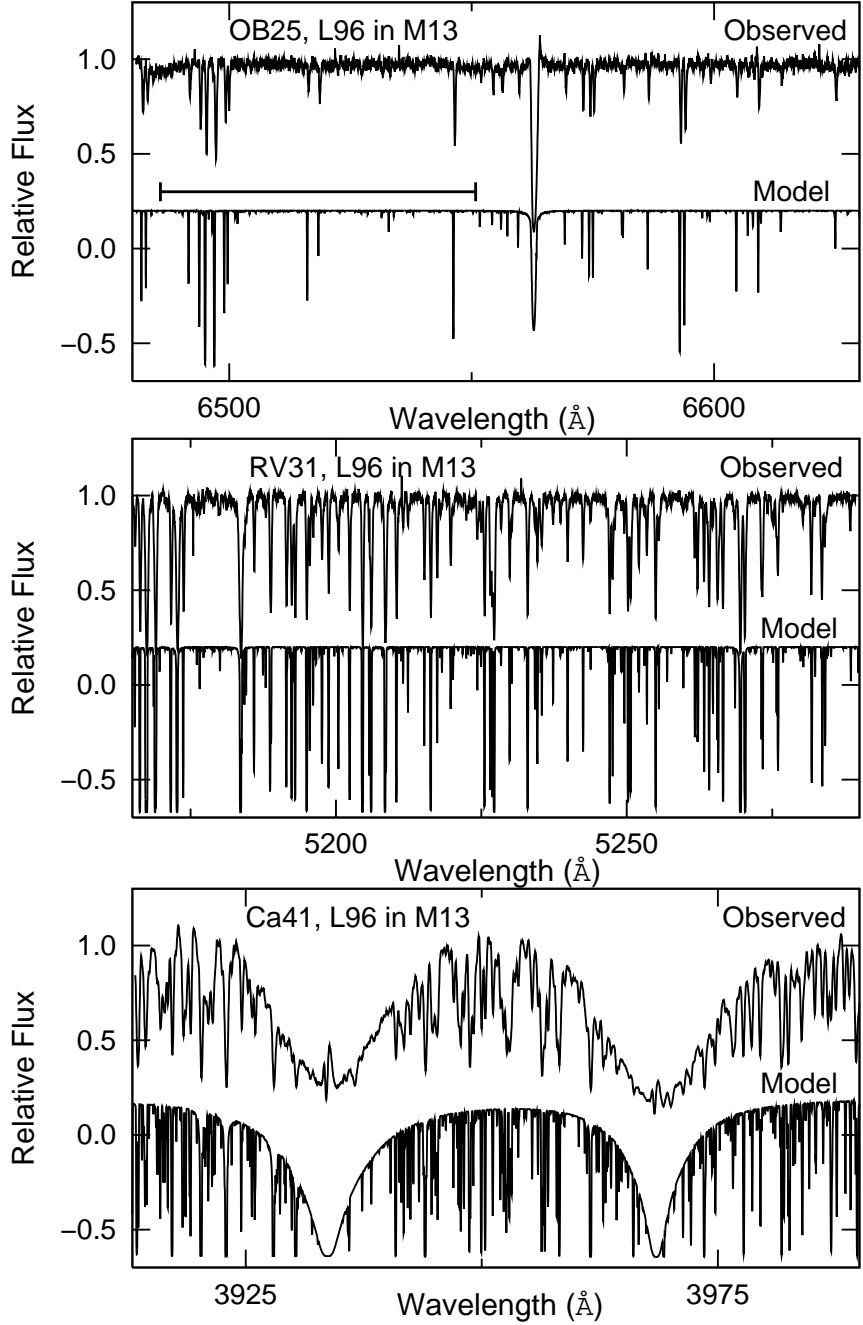}
\caption{Continuum normalized spectra of a sample star (L96) in M13 showing H$\alpha$, RV31, and \ion{Ca}{2} H$\&$K spectra 
after all reductions. Upper spectrum is the observed one, lower spectrum is the model synthesis of a star 
(Coelho et al. 2005, using Kurucz models) with the highest 
amplitude of the cross-correlation function from the H$\alpha$ region. 
The cross-correlation region used in the OB25 filter is marked in the spectrum and chosen to avoid H$\alpha$.}
\end{figure}

\clearpage

\begin{figure}  
\includegraphics[width=4.5in,angle=270]{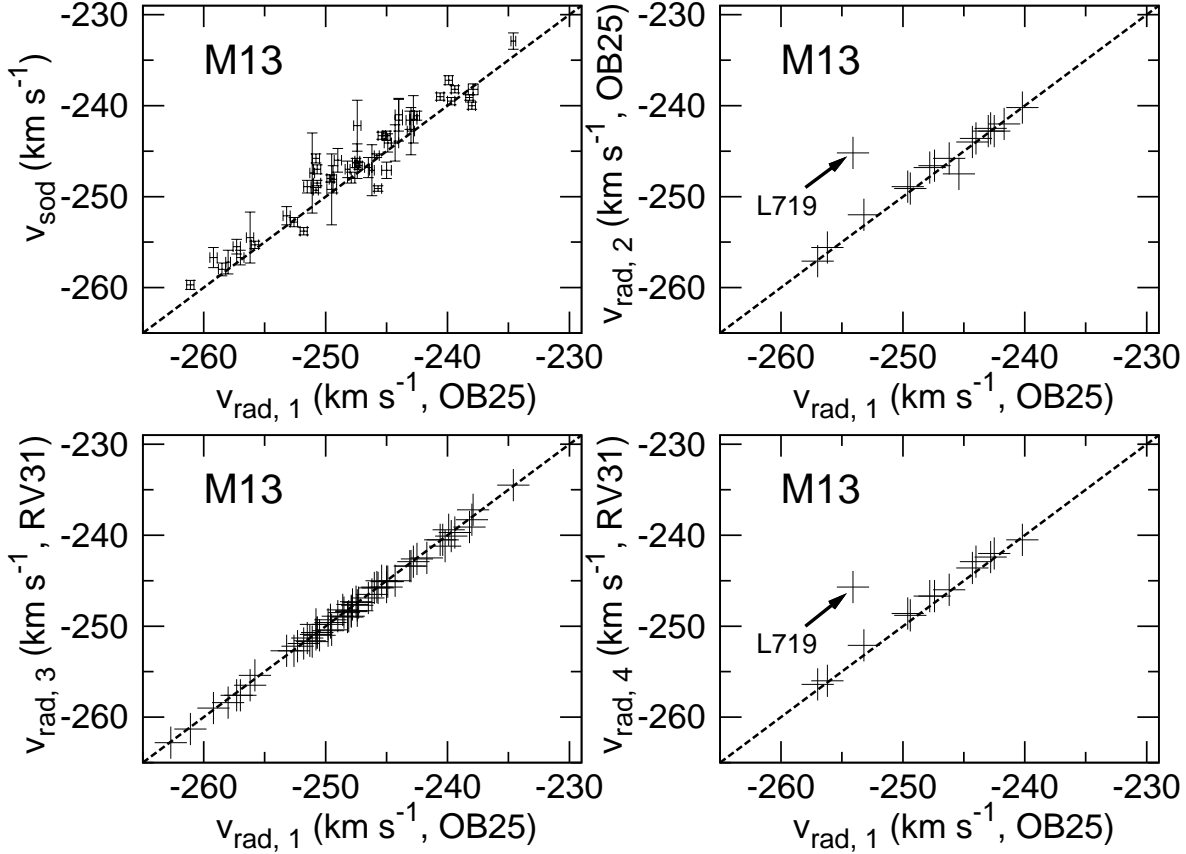}
\caption{Top left: Radial velocities measured with the H$\alpha$ filter (OB25) in this paper on 2006 March 14 
($v_{rad,1}$) compared to the same stars observed by \citet{soderberg01} ($v_{sod}$). There is a slight offset 
($1.1 \ \pm 0.5$~km~s$^{-1}$) between all observations taken in 2006 and observations for the same stars from 
\citet{soderberg01}. Top right: Radial velocities measured in this paper on 2006 March 14 ($v_{rad,1}$)
for the same stars observed on 2006 May 10 with the H$\alpha$ filter ($v_{rad,2}$). 
Lower left: Radial velocity measured with Hectochelle for the same stars observed on 2006 
March 14 ($v_{rad,1}$) compared to the observations with the RV31 filter on 2006 March 16 ($v_{rad,3}$). Lower right: 
Radial velocities for the same stars measured with Hectochelle on 2006 March 14 ($v_{rad,1}$) compared to 
the observations with the RV31 filter on 2006 May 10 ($v_{rad,4}$). 
The dashed line marks a 1:1 relation in all panels.  
The error of our measurements was generally smaller than the symbols used in the figure. The anomalous star in
M13, L719, lies between the AGB and RGB, and the large velocity change may indicate binarity.}
\end{figure}

\clearpage

\begin{figure}
\includegraphics[width=4.5in,angle=270]{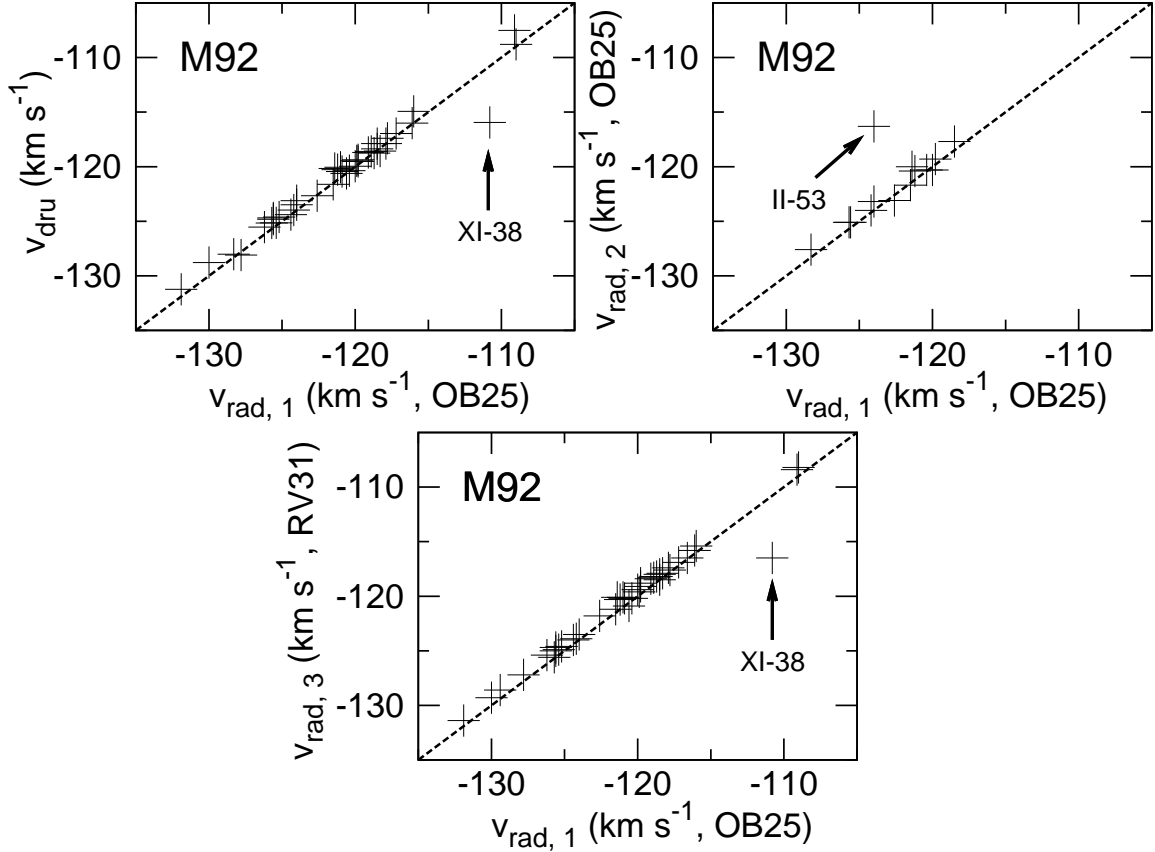}
\caption{Top left: Radial velocities measured with the H$\alpha$ filter in this paper on 2006 May 7 ($v_{rad,1}$) 
compared to the same stars observed by \citet{drukier01} ($v_{dru}$). 
Top right: Radial velocities measured in this paper on 2006 May 7 ($v_{rad,1}$)
for the same stars observed on 2006 May 9 with the H$\alpha$ filter ($v_{rad,2}$). 
Center: Radial velocity measured with Hectochelle for the same stars observed on 2006 
May 7 ($v_{rad,1}$) compared to the observations with the RV31 filter on the same day ($v_{rad,3}$). 
There is no offset larger than the error of measurements between any observations. 
The dashed line marks a 1:1 relation in all panels. There is good agreement between all
observations taken in May and observations for the same stars from \citet{drukier01}. The error of our measurements 
was generally smaller than the symbol we used in the figure. For
discussion  of the two outlier stars see Section 4.1.}
\end{figure}

\clearpage

\begin{figure}  
\includegraphics[width=4.5in,angle=270]{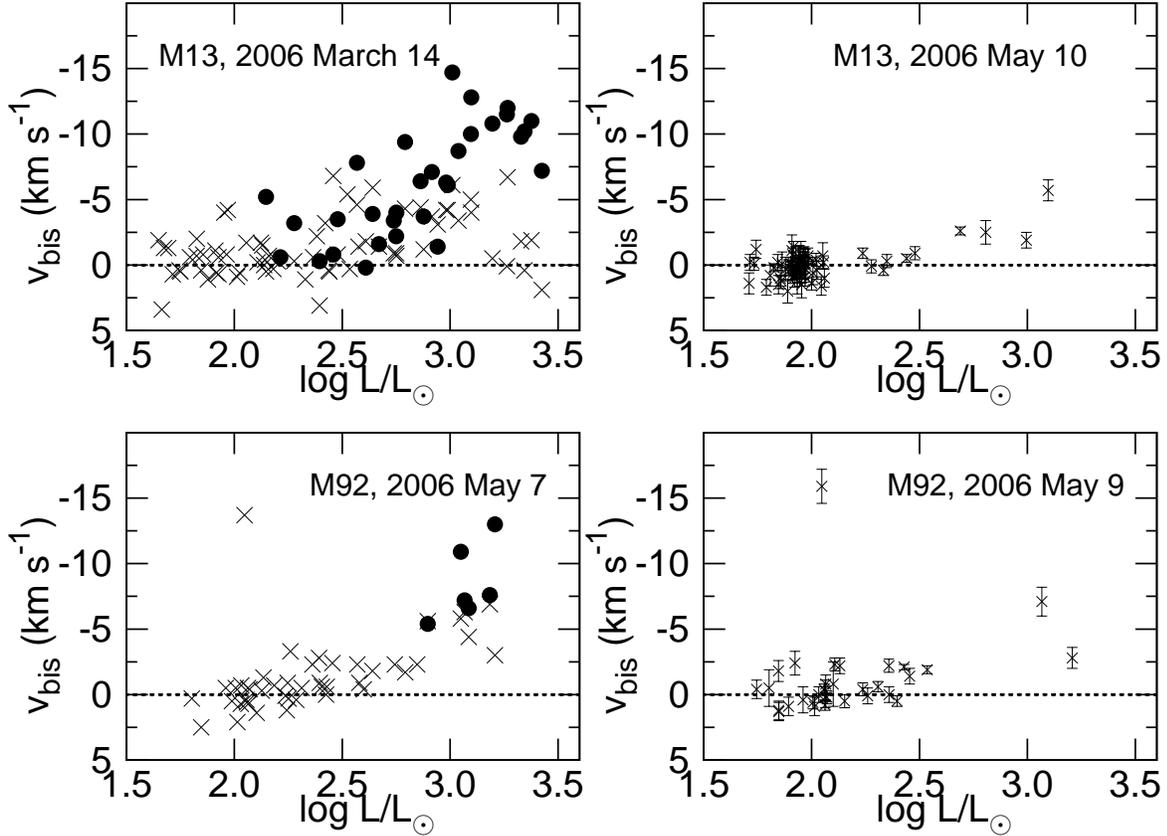}
\caption{The velocity difference ($v_{bis}$) between the top and the bottom of the
bisector of H$\alpha$ ($\times$) and the coreshift of the \ion{Ca}{2}
K central reversal 
absorption (filled circle) as a 
function of luminosity. All Ca observations from different days are
plotted together 
on the left side panels. 
Negative values indicate a blueshifted core (outward motion), positive
values denote 
a red shifted core (inward motion). The
error bars in figures on the left side were eliminated to display the
differences 
between H$\alpha$ and \ion{Ca}{2} K.
A predominant outward motion sets in near $log
L/L_{\odot}~\approx$~2.5 in both 
clusters and increases in velocity 
towards higher luminosity. The velocity of the \ion{Ca}{2} K central
reversal formed higher in the chromosphere than the H$\alpha$ core, 
is generally larger than the bisector velocity of H$\alpha$ at the
same luminosity, 
indicating that the expansion velocity increases with height in the chromosphere.}
\end{figure}

\clearpage

\begin{figure}  
\includegraphics[width=4.5in,angle=270]{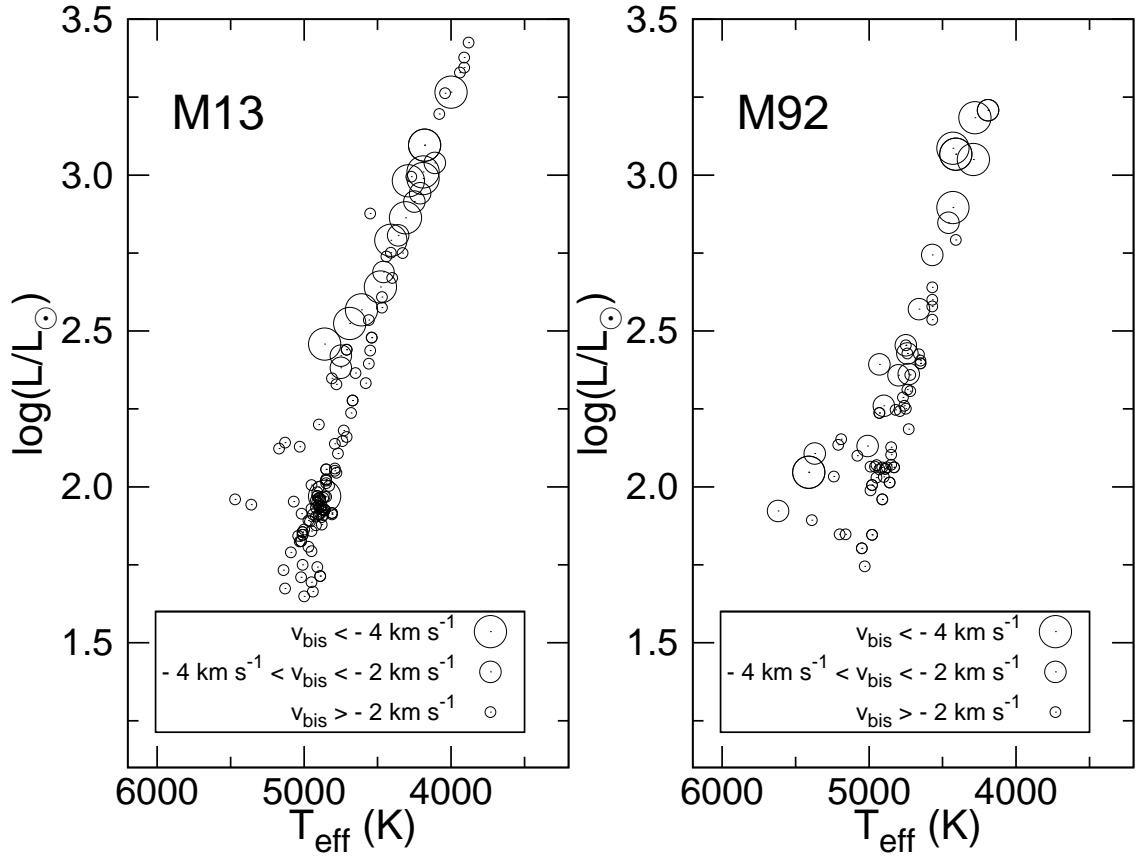}
\caption{Temperature-luminosity diagram for all stars observed in both clusters, where the size of the circle 
indicates the velocity of the H$\alpha$ bisector asymmetry. Big circle: $v_{bis} 
<-4$~km~s$^{-1}$, medium circle: $-4$ km s$^{-1} < v_{bis} < -2$~km~s$^{-1}$, small circle: 
$v_{bis} > -2$~km~s$^{-1}$. Concentric circles indicate multiple observations of the same star.}
\end{figure}

\clearpage

\begin{figure}   
\includegraphics[width=5in,angle=0]{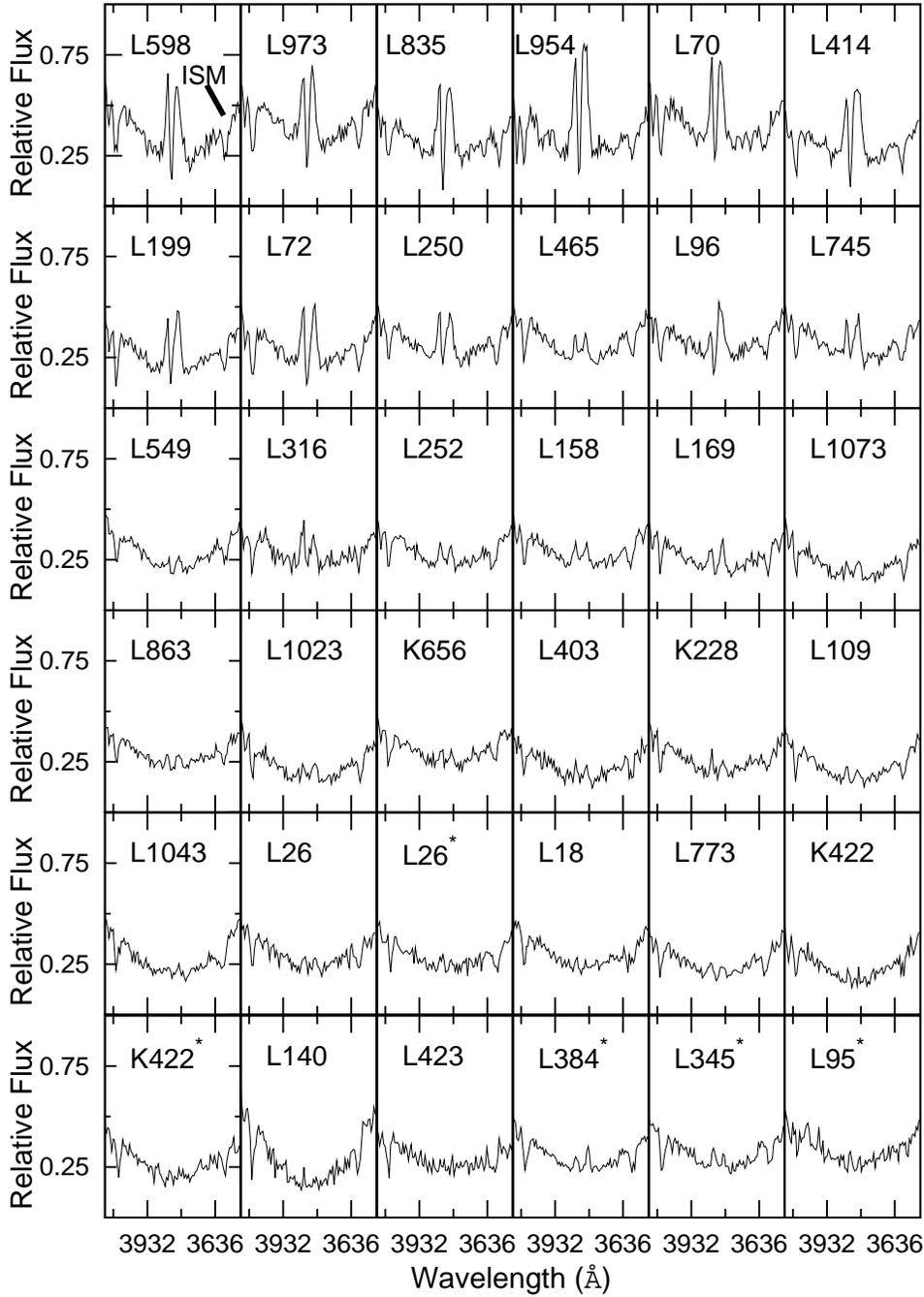}
\caption{Spectra of the brightest red giants in M13 which showed emission in \ion{Ca}{2}~K on 2006 March 14. 
The spectra are smoothed by 3. The spectra are arranged in order of decreasing brightness; the brightest 
is at the top left and the stars become fainter from left to right for a single date. The object names marked by stars
were observed on 2006 May 10. The wavelength scale is corrected for heliocentric 
velocity. The line marked ISM in the spectrum of L598 denotes
absorption by the interstellar medium and
can be recognized in the other spectra.}
\end{figure}

\clearpage

\begin{figure}
\includegraphics[width=5in,angle=0]{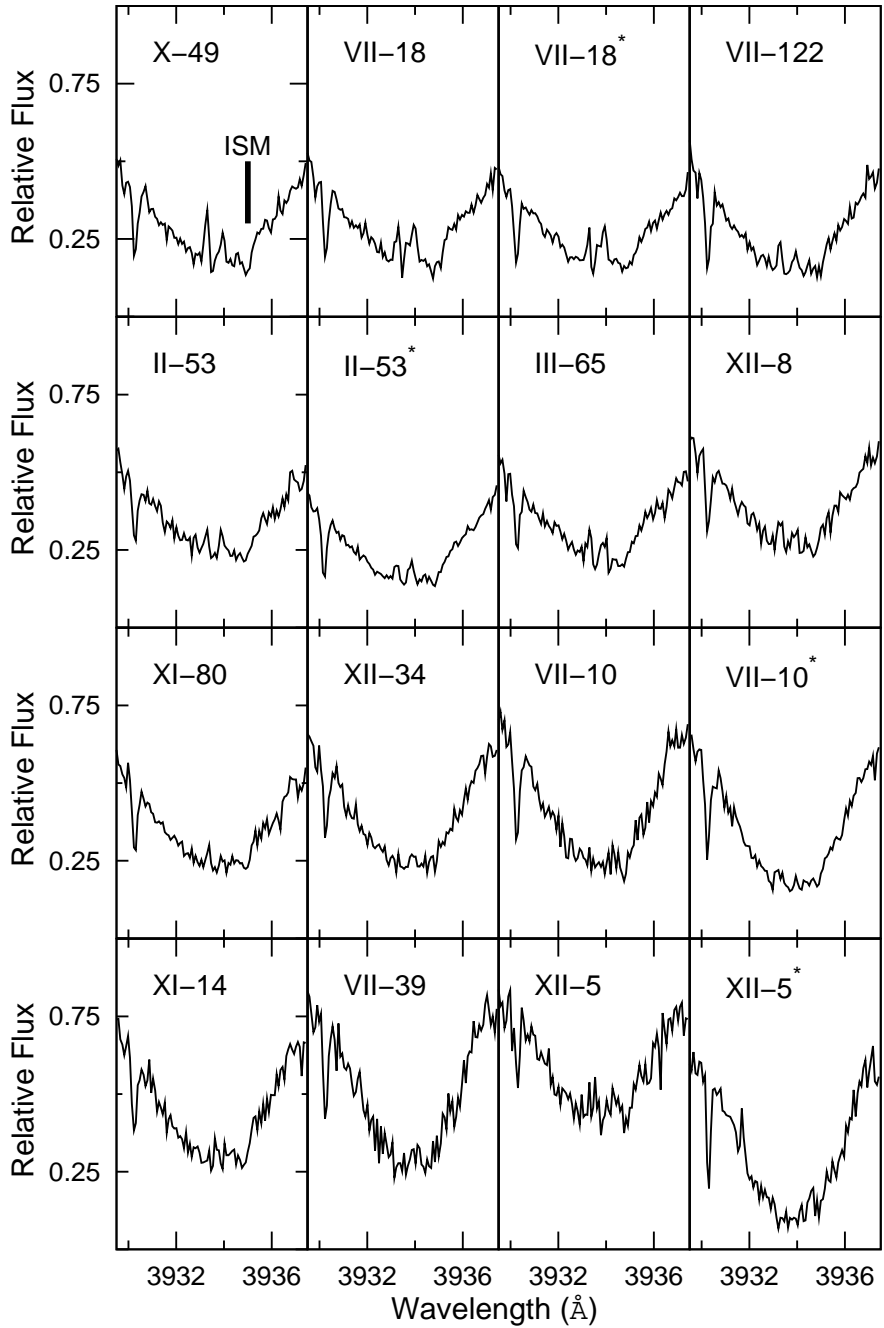}
\caption{Spectra of the brightest red giants in M92 which showed emission in \ion{Ca}{2}~K on 2006 May 7. 
The object names marked by stars were observed on 2006 May
9. Observations obtained on 2006 May 7 have generally lower S/N due to bad sky 
conditions and resulted in a higher \ion{Ca}{2}~K core of XII-5, the faintest star
in our sample. Additional explanation can be found in  the caption of Figure 11.}
\end{figure}

\clearpage

\begin{figure}
\includegraphics[width=4.5in,angle=270]{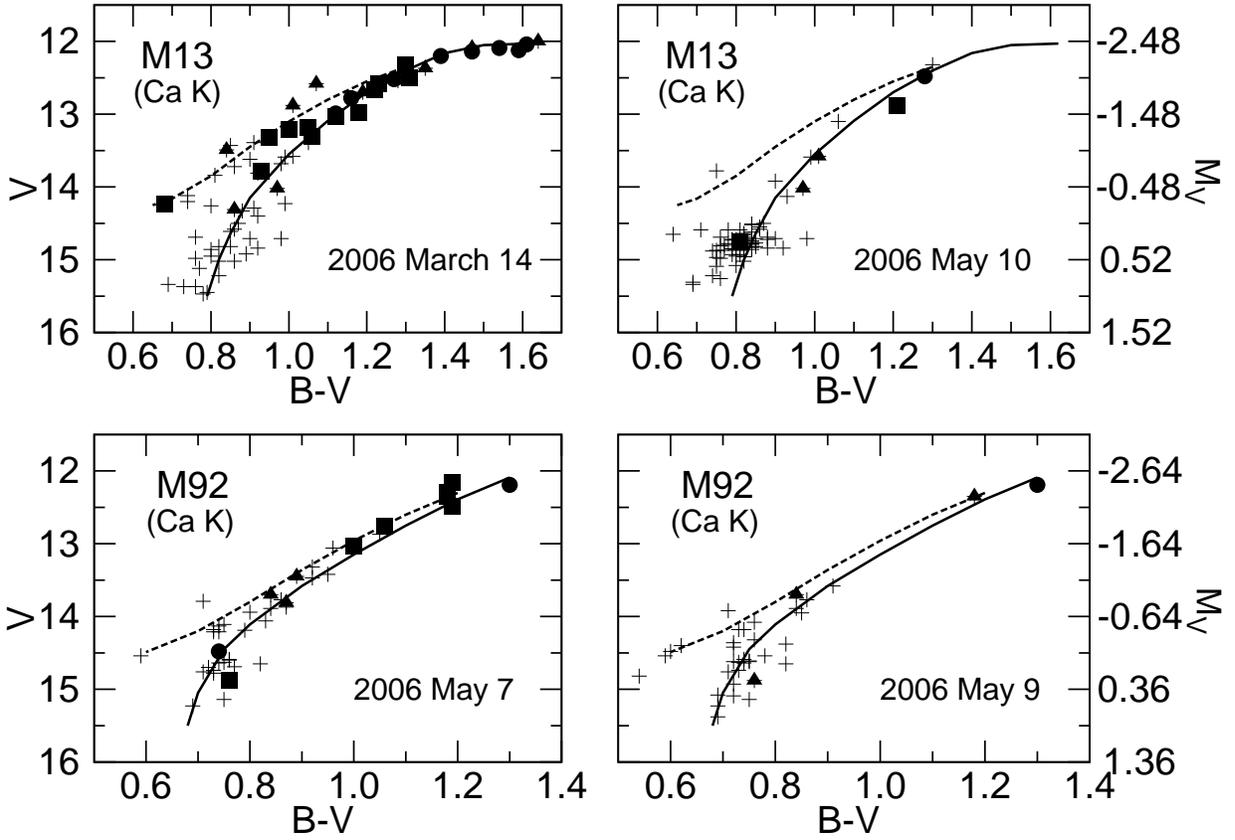}
\caption{Color-magnitude diagrams for all M13 and M92 stars observed. 
Stars with \ion{Ca}{2}~K emission and with B$<$R (indicating outflow) are marked with circles; 
stars with B$>$R emission wings (suggests inflow) are denoted by squares and stars with 
B$\approx$R are marked with triangles.
The solid line shows the fiducial curve of the RGB; dashed lines show the fiducial curve of the 
AGB for M13 and M92 from observations of \citet{sandage01}.}
\end{figure}

\clearpage

\begin{figure}
\includegraphics[width=4.5in,angle=270]{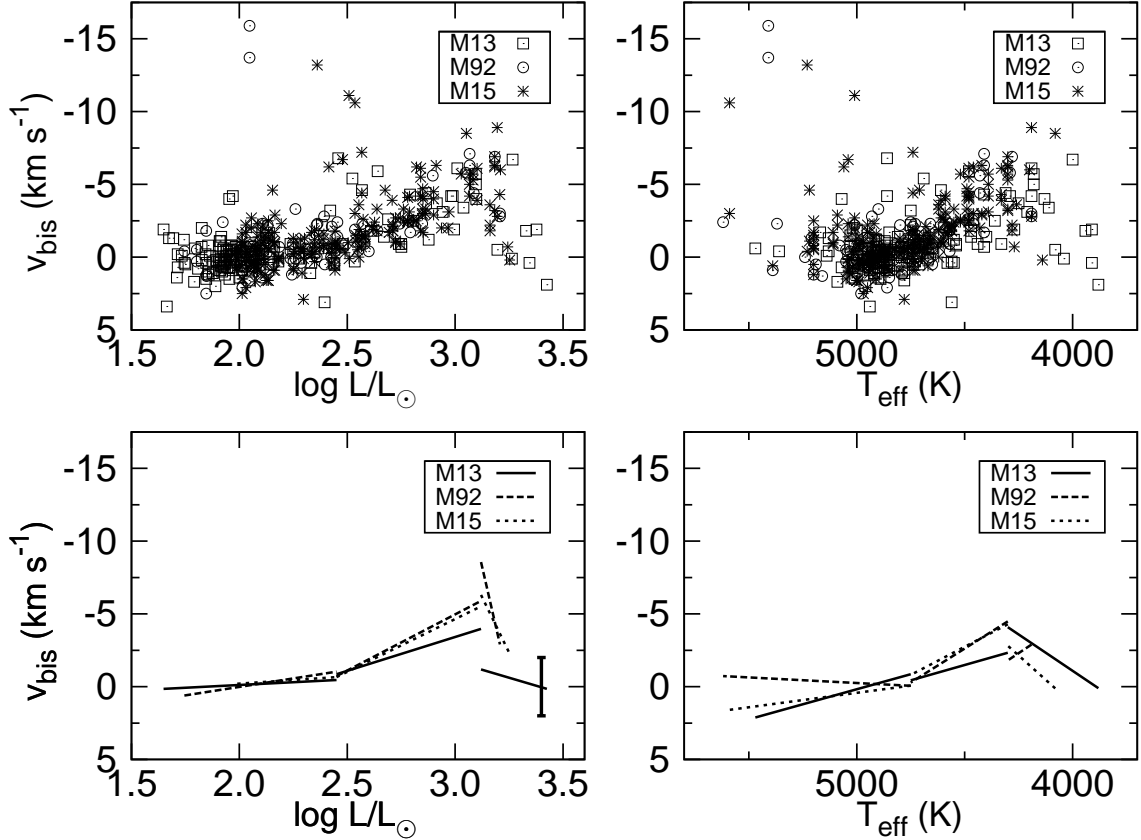}
\caption{{\it Top left and right:} The bisector velocity ($v_{bis}$) of H$\alpha$ for 
all observations in M13, M15 and M92 as a 
function of luminosity and effective temperature. Negative values indicate 
outflow. {\it Lower left and right:} To fit the results with a linear function, luminosity
  and effective temperature were divided into three
different regions: the bottom of the RGB [log~$(L/L_{\odot})=1.6-2.5$, 
T$_{eff}$=4750$-$5700~K], where $v_{bis}$ was close to 
zero ~km~s$^{-1}$; the RGB stars [log~$(L/L_{\odot})=2.5-3.1$, 
T$_{eff}$=4300$-$4750~K], where $v_{bis}$ shows a significant
increase; and the top of the RGB [log~$(L/L_{\odot})=3.1-3.5$, 
T$_{eff}$=3800$-$4300~K], where $v_{bis}$ generally shows smaller
values than in the middle of the RGB. The stars with high velocity near 
log~$(L/L_{\odot})=2.2-2.4$ were omitted from the fit. Error bar of the top of the RGB fit for M13 is displayed 
in the lower left panel (for details see Section 6.3). The errors of the fits span between $\sim$ 1 and 
2.5~km~s$^{-1}$ for each fit.}
\end{figure}

\clearpage

\begin{deluxetable}{lcccccccccc}
\tabletypesize{\scriptsize}
\tablecaption{Photometric Data of Observed Cluster Members in M13}
\tablewidth{0pt}
\tablehead{
\colhead{ID No. \tablenotemark{a}}           & \colhead{RA(2000) \tablenotemark{b} }    &
\colhead{Dec(2000) \tablenotemark{b} }          & \colhead{B}  &
\colhead{V}          & \colhead{J}    &
\colhead{H}  & \colhead{K}  &
\colhead{B$-$V} & \colhead{V$-$K} & \colhead{Obs. \tablenotemark{c} }
}
\startdata
III-65 \tablenotemark{d} & 16 41 39.091 & +36 23 51.40 & 16.04 & 15.22 & 13.690 & 13.220 & 13.071 & 0.82 & 2.15 & 1,3,5 \\
K188 \tablenotemark{e} & 16 40 42.982 & +36 27 41.88 & 14.44 & 13.39 & 11.410 & 10.784 & 10.704 & 1.05 & 2.69 & 1,3,5 \\ 
K210 \tablenotemark{e} & 16 40 56.378 & +36 22 18.51 & 15.21 & 14.33 & 12.580 & 12.028 & 11.966 & 0.88 & 2.36 & 1,3,5 \\ 
K220 \tablenotemark{e} & 16 41 02.608 & +36 26 15.81 & 15.74 & 14.98 & 13.353 & 12.899 & 12.833 & 0.76 & 2.15 & 1,2,4,5,6 \\ 
K223 \tablenotemark{e} & 16 41 05.075 & +36 28 20.85 & 15.56 & 14.71 & 13.093 & 12.565 & 12.494 & 0.85 & 2.22 & 2,4,6 \\
\enddata
\tablenotetext{a}{\citet{ludendorff01} is the identification for the majority of the stars denoted by L.}
\tablenotetext{b}{2MASS coordinates \citep{skrutskie01}}
\tablenotetext{c}{Observations: 1: 2006 March 14 (OB25), 2: 2006 May 10 (OB25), 3: 2006 March 16 (RV31), 
4: 2006 May 10 (RV31), 5: 2006 March 16 (Ca41), 6: 2006 May 10 (Ca41).}
\tablenotetext{d}{\citet{arp01}}
\tablenotetext{e}{\citet{kadla01}}
\tablecomments{The visual photometry is taken from \citet{cudworth02}, 
J,H,K photometry is taken from the 2MASS Catalog \citep{skrutskie01}. 
This table is available in its entirety in a machine-readable form
in the on-line journal.  A portion is shown here for guidance regarding
its form and content.}
\end{deluxetable}


\begin{deluxetable}{lcccccccccc}
\tabletypesize{\scriptsize}
\tablecaption{Photometric Data of Observed Cluster Members in M92}
\tablewidth{0pt}
\tablehead{
\colhead{ID No. \tablenotemark{a}}           & \colhead{RA(2000) \tablenotemark{b} }    &
\colhead{Dec(2000) \tablenotemark{b} }          & \colhead{B}  &
\colhead{V}          & \colhead{J}    &
\colhead{H}  & \colhead{K}  &
\colhead{B$-$V} & \colhead{V$-$K} & \colhead{Obs. \tablenotemark{c} }
}
\startdata
I-14 	& 17 17 28.77 & 43 10 02.8 & 15.47 & 14.74 & 13.155 & 12.644 & 12.592 & 0.73 & 2.148 & 1,2,3,4,5 \\
I-40 	& 17 17 22.68 & 43 08 50.5 & 15.51 & 14.78 & 13.258 & 12.777 & 12.640 & 0.73 & 2.140 & 1,3,4 \\
I-67 	& 17 17 21.24 & 43 08 27.0 & 14.24 & 13.32 & 11.406 & 10.870 & 10.766 & 0.92 & 2.554 & 1,3,4 \\
I-68 	& 17 17 21.73 & 43 08 15.8 & 15.36 & 14.61 & 13.243 & 12.825 & 12.661 & 0.75 & 1.949 & 2,5 \\
II-6 	& 17 17 50.37 & 43 13 46.0 & 15.89 & 15.14 & 13.541 & 13.002 & 12.992 & 0.75 & 2.148 & 1,2,3,4,5 \\
\enddata
\tablenotetext{a}{\citet{sandage02}}
\tablenotetext{b}{2MASS coordinates \citep{skrutskie01}}
\tablenotetext{c}{Observations: 1: 2006 May 7 (OB25), 2: 2006 May 9 (OB25), 3: 2006 May 7 (RV31), 4: 2006 May 8 (Ca41), 
5: 2006 May 9 (Ca41).}
\tablecomments{The visual photometry is taken from \citet{cudworth01}, 
J,H,K photometry is taken from the 2MASS Catalog \citep{skrutskie01}.
This table is available in its entirety in a machine-readable form
in the on-line journal.  A portion is shown here for guidance regarding
its form and content.}

\end{deluxetable}

\clearpage

\begin{deluxetable}{lcccc}
\tabletypesize{\scriptsize}
\tablecaption{Hectochelle Observations of M13 and M92}
\tablewidth{0pt}
\tablehead{
\colhead{Date}           & \colhead{Total exp.}      &
\colhead{Wavelength} & \colhead{Filter Name} & \colhead{Number of }    \\
\colhead{(UT)} & \colhead{(s)} & \colhead{(\AA)} & \colhead{} & \colhead{Observed Stars} 
}
\startdata
2006 March 14 (M13, Field 1) & $3 \times 2400$ & 6475$-$6630 & OB25 & 70 \\
2006 March 16 (M13, Field 1) & $3 \times 2400$ & 3910$-$3990 & Ca41 & 70 \\
2006 March 16 (M13, Field 1) & $1 \times 2400$ & 5150$-$5300 & RV31 & 65 \\
2006 May 10 (M13, Field 2) & $3 \times 2400$ & 6475$-$6630 & OB25 & 70 \\
2006 May 10 (M13, Field 2) & $3 \times 2400$ & 3910$-$3990 & Ca41 & 63 \\
2006 May 10 (M13, Field 2) & $1 \times 2400$ & 5150$-$5300 & RV31 & 65 \\
2006 May 7 (M92, Field 1) & $3 \times 2400$ & 6475$-$6630 & OB25 & 42 \\
2006 May 7 (M92, Field 1) & $3 \times 1800$ & 5150$-$5300 & RV31 & 40 \\
2006 May 8 (M92, Field 1) & $3 \times 2400$ & 3910$-$3990 & Ca41 & 41 \\
2006 May 9 (M92, Field 2) & $3 \times 1800$ & 6475$-$6630 & OB25 & 36 \\
2006 May 9 (M92, Field 2) & $3 \times 2400$ & 3910$-$3990 & Ca41 & 36 \\
\enddata
\end{deluxetable}

\clearpage

\begin{deluxetable}{lcclcc}
\tabletypesize{\scriptsize}
\tablecaption{B/R ratio of H$\alpha$ Line for Stars with Emission Wings}
\tablewidth{0pt}
\tablehead{
\colhead{} & \colhead{M13} & \colhead{} & \colhead{} & \colhead{M92} & \colhead{} \\
\colhead{} & \colhead{$B/R$} & \colhead{$B/R$} & \colhead{} & \colhead{$B/R$} & \colhead{$B/R$} \\
\colhead{ID No.}           & \colhead{2006 March 14}      &
\colhead{2006 May 10} & \colhead{ID No.} & \colhead{2006 May 7} & \colhead{2006 May 9}
}
\startdata
L70 & $<1$ & \nodata & II-53 & $>1$ & $>1$ \\
L72 & $<1$ & $>1$ & III-65 & $>1$ & \nodata \\
L96 & $<1$ & \nodata & IV-94 & $>1$ & \nodata \\
L158 & $>1$ & \nodata & VII-18 & $<1$ & $<1$ \\
L169 & $>1$ & \nodata & VII-122 & $>1$ & \nodata \\
L199 & $<1$ & \nodata & IX-12 & $>1$ & $>1$ \\
L250 & $>1$ & \nodata & X-49 & $>1$ & \nodata \\
L252 & $>1$ & \nodata & XII-8 & $>1$ & \nodata \\
L316 & $>1$ & \nodata & & & \\ 
L345 & \nodata & $>1$ & & & \\ 
L414 & $<1$ & \nodata & & & \\
L465 & $>1$ & \nodata & & & \\
L598 & $<1$ & \nodata & & & \\
L719 & $<1$ & no emission & & & \\
L745 & $>1$ & \nodata & & & \\
L835 & $<1$ & \nodata & & & \\
L954 & $<1$ & \nodata & & & \\
L973 & $>1$ & \nodata & & & \\
L1073 & $>1$ & \nodata & & & \\
\enddata
\tablecomments{The parameter B/R is the intensity 
ratio of Blue (short wavelength) and Red (long wavelength) emission peaks. The symbol \nodata indicates the star was not
observed. If B/R ratio is $>1$ the emission wings indicate inflow, if B/R ratio is $<1$ the emission wings indicate 
outflow.}
\end{deluxetable}

\clearpage

\begin{deluxetable}{lccccccr}
\tabletypesize{\scriptsize}
\tablecaption{Physical Parameters of Cluster Members in M13}
\tablewidth{0pt}
\tablehead{
\colhead{ID No.}           & \colhead{$M_{V}$}      &
\colhead{$(B-V)_0$}          & \colhead{$(V-K)_0$}  &
\colhead{P \tablenotemark{a}}          & \colhead{T$_{eff}$} & \colhead{log~$L/L_{\odot}$} & \colhead{$R/R_{\odot}$}\\
\colhead{} & \colhead{} & \colhead{} & \colhead{} & \colhead{} & \colhead{(K)} & \colhead{} & \colhead{} 
}
\startdata
III-65 & +0.74 & 0.80 & 2.095 & 99 & 	5010 & 1.750 &   9.7  \\
K188 & $-1.09$ & 1.03 & 2.635 & 99 & 	4470 & 2.575 &   31.5  \\
K210 & $-0.15$ & 0.86 & 2.305 & 99 & 	4790  & 2.138 &  16.6   \\
K220 & +0.50 & 0.74 & 2.095 & 99 & 	5010 & 1.846 &   10.8  \\
K223 & +0.23 & 0.83 & 2.165 & 99 & 	4900 & 1.964 &   13.0  \\
\enddata
\tablenotetext{a}{Membership probability from proper motion
  observations \citep{cudworth02}.}
\tablecomments{This table is available in its entirety in a machine-readable form
in the on-line journal.  A portion is shown here for guidance regarding
its form and content.}
\end{deluxetable}


\begin{deluxetable}{lccccccr}
\tabletypesize{\scriptsize}
\tablecaption{Physical Parameters of Cluster Members in M92}
\tablewidth{0pt}
\tablehead{
\colhead{ID No.}           & \colhead{$M_{V}$}      &
\colhead{$(B-V)_0$}          & \colhead{$(V-K)_0$}  &
\colhead{P \tablenotemark{a}}          & \colhead{T$_{eff}$} & \colhead{log~$L/L_{\odot}$} & \colhead{$R/R_{\odot}$}\\
\colhead{} & \colhead{} & \colhead{} & \colhead{} & \colhead{} & \colhead{(K)} & \colhead{} & \colhead{} 
}
\startdata
I-14 &+0.10  & 0.71 & 2.093 & 99 &    4980 & 2.006 & 13.5 \\
I-40 &+0.14  & 0.71 & 2.085 & 99 &    4990 & 1.989 & 13.2 \\
I-67 &$-1.32$ & 0.90 & 2.499 & 99 &    4570 & 2.640 & 33.4 \\
I-68 &$-0.03$ & 0.73 & 1.894 & 99 &    5240 & 2.033 & 12.6 \\
II-6 &+0.50 & 0.73 & 2.093 & 99 &      4980 & 1.846 & 11.3 \\
\enddata
\tablenotetext{a}{Membership probability from proper motion observations \citep{cudworth01}.}
\tablecomments{This table is available in its entirety in a machine-readable form
in the on-line journal.  A portion is shown here for guidance regarding
its form and content.}
\end{deluxetable}

\clearpage

\begin{deluxetable}{lcccccc}
\tabletypesize{\scriptsize}
\tablecaption{Radial and H$\alpha$ Bisector Velocity of Observed Stars in M13}
\tablewidth{0pt}
\tablehead{
\colhead{ID No.} &
\colhead{$v_{rad, 1}$ \tablenotemark{a}}         & 
\colhead{$v_{rad, 2}$ \tablenotemark{a}}        & 
\colhead{$v_{rad, 3}$ \tablenotemark{a}}        & 
\colhead{$v_{rad, 4}$ \tablenotemark{a}}  &
\colhead{$v_{bis, 1}$ \tablenotemark{b}}	  & 
\colhead{$v_{bis, 2}$ \tablenotemark{b}}      \\
\colhead{} & \colhead{\tiny{(km \ s$^{-1}$)}} & \colhead{\tiny{(km \ s$^{-1}$)}} & 
\colhead{\tiny{(km \ s$^{-1}$)}} & 
\colhead{\tiny{(km \ s$^{-1}$)}} &
\colhead{\tiny{(km \ s$^{-1}$)}} & 
\colhead{\tiny{(km \ s$^{-1}$)}}
}
\startdata
III-65 & $-246.0 \ \pm$ 0.4 & \nodata  &  $-245.9 \ \pm$ 0.2 & \nodata 			       & $+0.5 \ \pm$ 2.5  & \nodata  	  \\				     
K188 &	$-244.9 \ \pm$ 0.3  & \nodata  & $-245.2 \ \pm$ 0.2  &\nodata	   		       & $-1.4 \ \pm$ 1.0  & \nodata   	 \\				      
K210 &	$-250.8 \ \pm$ 0.2  & \nodata  & $-249.9 \ \pm$ 0.3  &\nodata	     		       & $-0.0 \ \pm$ 0.6  & \nodata   	 \\			      
K220 &	$-244.0 \ \pm$ 0.3  & $-243.6 \ \pm$ 0.2  & \nodata &	$-242.9 \ \pm$ 0.2   	       & $+0.4 \ \pm$ 1.4  & $+1.5 \ \pm$ 0.7   \\				    
K223 &	\nodata & $-242.6 \ \pm$ 0.2  & \nodata  &  $-242.0 \ \pm$ 0.2	      		       & \nodata  & $+0.2 \ \pm$ 0.9   	 \\	
\enddata 
\tablenotetext{a}{Observations: 1: 2006 March 14 (H$\alpha$), 2: 2006 May 10 (H$\alpha$), 
3: 2006 March 16 (RV31), 4: 2006 May 10 (RV31).}
\tablenotetext{b}{Observations: 1: 2006 March 14; 2: 2006 May 10.}
\tablecomments{This table is available in its entirety in a machine-readable form
in the on-line journal.  A portion is shown here for guidance regarding
its form and content.}
\end{deluxetable}

	 
\begin{deluxetable}{lccccc}
\tabletypesize{\scriptsize}
\tablecaption{Radial and H$\alpha$ Bisector Velocity of Observed Stars in M92}
\tablewidth{0pt}
\tablehead{
\colhead{ID No.} &
\colhead{$v_{rad, 1}$ \tablenotemark{a}}         & 
\colhead{$v_{rad, 2}$ \tablenotemark{a}}        & 
\colhead{$v_{rad, 3}$ \tablenotemark{a}}        & 
\colhead{$v_{bis, 1}$ \tablenotemark{b}}	  & 
\colhead{$v_{bis, 2}$ \tablenotemark{b}}      \\
\colhead{} & \colhead{\tiny{(km \ s$^{-1}$)}} & \colhead{\tiny{(km \ s$^{-1}$)}} & 
\colhead{\tiny{(km \ s$^{-1}$)}} & 
\colhead{\tiny{(km \ s$^{-1}$)}} & \colhead{\tiny{(km \ s$^{-1}$)}} 
}
\startdata
I-14 & $-124.2 \ \pm$ 0.3 & $-124.0 \ \pm$ 0.2 & $-123.9 \ \pm$ 0.2 &   	$-0.5 \ \pm$ 1.2  & $+0.6 \ \pm$ 0.5  \\
I-40 & $-125.2 \ \pm$ 0.2 & \nodata  & $-124.6 \ \pm$ 0.2 &    	      		$+0.5 \ \pm$ 0.5  & \nodata            \\
I-67 & $-120.9 \ \pm$ 0.2 & \nodata  & $-120.2 \ \pm$ 0.2 &    	      		$-1.8 \ \pm$ 0.4 &  \nodata 	     \\
I-68 &	\nodata & $-129.7 \ \pm$ 0.2  & \nodata &     	      			\nodata  & $+0.0 \ \pm$ 0.6          \\
II-6 & $-122.6 \ \pm$ 0.3  & $-123.1 \ \pm$ 0.3 & $-121.8 \ \pm$ 0.2 &  	$+2.5 \ \pm$ 0.7  & $-1.8 \ \pm$ 0.8  \\  
\enddata 
\tablenotetext{a}{Observations: 1: 2006 May 7 (OB25), 2: 2006 May 9 (OB25), 3: 2006 May 7 (RV31)}
\tablenotetext{a}{Observations: 1: 2006 March 14; 2: 2006 May 10.}
\tablecomments{This table is available in its entirety in a machine-readable form
in the on-line journal.  A portion is shown here for guidance regarding
its form and content.}
\end{deluxetable}

\clearpage

\begin{deluxetable}{lcclcc}
\tabletypesize{\scriptsize}
\tablecaption{B/R ratio of \ion{Ca}{2}~K Line for Stars Showing Emission in M13 and M92}
\tablewidth{0pt}
\tablehead{
\colhead{} & \colhead{M13} & \colhead{} & \colhead{} & \colhead{M92} & \colhead{} \\
\colhead{} & \colhead{$B/R$} & \colhead{$B/R$} & \colhead{} & \colhead{$B/R$} & \colhead{$B/R$} \\
\colhead{ID No.} & \colhead{2006 March 16} & \colhead{2006 May 10} & \colhead{ID No.} & \colhead{2006 May 8} &
\colhead{2006 May 9}
}
\startdata
K228 & $>1$ & \nodata  	    & II-53	& $>1$ & $\approx 1$	 \\
K422 & $\approx 1$ &  $\approx 1$	    & III-65	& $>1$ & \nodata	 \\
K656 & $>1$ & \nodata  	    & VII-10	& $\approx 1$ & $\approx 1$	 \\
L18 & $>1$ & \nodata   	    & VII-18	& $<1$ & $<1$	 \\
L26 & $>1$  & $\approx 1$  	    & VII-39	& $<1$ & \nodata 	 \\
L70 & $<1$ &  \nodata  	    & VII-122	& $>1$ & \nodata	 \\
L72 & $<1$ &  \nodata  	    & X-49	& $>1$ & \nodata	 \\
L95 & \nodata  & $>1$  	    & XI-14 	& $\approx 1$ & \nodata 	 \\
L96 & $<1$ & \nodata   	    & XI-80	& $>1$ & \nodata	   \\
L109 & $>1$ & \nodata  	    & XII-5	& $>1$ & $\approx 1$	    \\
L140 & $>1$ & \nodata       & XII-8	& $>1$ & \nodata	   \\
L158 & $\approx 1$ & \nodata       & XII-34	& $\approx 1$ & \nodata	   \\
L169 & $<1$ & \nodata       & & &  \\
L199 & $<1$ & \nodata       & & &  \\
L250 & $\approx 1$ & \nodata       & & &  \\
L252 & $>1$ & \nodata       & & &  \\
L316 & $>1$ & \nodata       & & &  \\
L345 & \nodata  & $>1$ 	    & & &  \\
L384 & \nodata  & $<1$ 	    & & &  \\
L403 & $>1$ & \nodata       & & &  \\
L414 & $<1$ & \nodata       & & &  \\
L423 & $\approx 1$ & \nodata       & & &  \\
L465 & $>1$ & \nodata       & & &  \\
L549 & $\approx 1$ & \nodata       & & &  \\
L598 & $\approx 1$ & \nodata       & & &  \\
L745 & $\approx 1$ & \nodata       & & &  \\
L773 & $>1$ & \nodata       & & &  \\
L835 & $\approx 1$ & \nodata       & & &  \\
L863 & $>1$ & \nodata 	    & & &  \\
L954 & $<1$ & \nodata       & & &  \\
L973 & $<1$ &  \nodata      & & &  \\
L1023 & $>1$ & \nodata      & & &  \\
L1043 & $\approx 1$ & \nodata      & & &  \\
L1073 & $\approx 1$ & \nodata      & & &  \\
\enddata		    
\tablecomments{The parameter $B/R$ is the intensity 
ratio of Blue (short wavelength) and Red (long wavelength) emission peaks.}
\end{deluxetable}

\clearpage		
	
\begin{deluxetable}{lcclcc}
\tabletypesize{\scriptsize}
\tablecaption{Relative Radial Velocity of \ion{Ca}{2}~K central absorption.}
\tablewidth{0pt}
\tablehead{
\colhead{} & \colhead{M13} & \colhead{} & \colhead{} & \colhead{M92} & \colhead{} \\
\colhead{ID No.} & \colhead{} & \colhead{$v_{rel}$} & \colhead{ID No.} & \colhead{} & \colhead{$v_{rel}$} \\
\colhead{} & \colhead{} & \colhead{\tiny{(km \ s$^{-1}$)}} & \colhead{} & \colhead{} & \colhead{\tiny{(km \ s$^{-1}$)}}
}
\startdata
K228 & & $+0.2 \ \pm$ 0.8 & II-53 & & $-7.2 \ \pm$ 0.9	  \\
K422 & & $-3.2 \ \pm$ 0.9  & III-65 & & $-10.9 \ \pm$ 0.9   \\
K656 & & $-2.2 \ \pm$ 0.7 & VII-18 & & $-13.0 \ \pm$ 0.4    \\
L18 & & $-0.3 \ \pm$ 0.8  & VII-122 & & $-6.6 \ \pm$ 0.8    \\
L26 & & $-3.5 \ \pm$ 0.9  & X-49 & & $-7.6 \ \pm$ 1.0	  \\
L70 & & $-10.2 \ \pm$ 0.5 & XII-8 & & $-5.4 \ \pm$ 0.8	  \\
L72 & & $-10.0 \ \pm$ 0.9 \\
L96 & & $-14.7 \ \pm$ 0.9 \\
L109 & & $-7.8 \ \pm$ 1.0 \\
L140 & & $-0.6 \ \pm$ 0.6 \\
L158 & & $-7.1 \ \pm$ 0.9 \\
L169 & & $-6.4 \ \pm$ 0.9 \\
L199 & & $-10.8 \ \pm$ 0.8 \\
L250 & & $-12.8 \ \pm$ 1.0 \\
L252 & & $-1.4 \ \pm$ 0.8 \\
L316 & & $-6.1 \ \pm$ 0.7 \\
L403 & & $-1.6 \ \pm$ 0.9 \\
L414 & & $-12.0 \ \pm$ 1.1 \\
L423 & & $-5.2 \ \pm$ 1.1 \\
L465 & & $-6.3 \ \pm$ 0.9 \\
L549 & & $-3.7 \ \pm$ 1.2  \\
L598 & & $-7.2 \ \pm$ 0.2 \\
L745 & & $-8.7 \ \pm$ 0.8 \\
L773 & & $-3.9 \ \pm$ 1.1 \\
L835 & & $-11.5 \ \pm$ 0.9 \\
L863 & & $-4.0 \ \pm$ 1.1 \\
L954 & & $-9.8 \ \pm$ 1.1 \\
L973 & & $-11.0 \ \pm$ 0.6 \\
L1023 & & $-3.4 \ \pm$ 1.1 \\
L1043 & & $-0.8 \ \pm$ 0.9 \\
L1073 & & $-9.4 \ \pm$ 1.0 \\
\enddata		    
\tablecomments{The table does not contain stars where the central absorption was not visible in the spectrum.}	
\end{deluxetable}		  

\clearpage		
			
\begin{deluxetable}{lccccccc}
\tabletypesize{\scriptsize}
\tablecaption{Characteristics of Emission in Four Clusters}
\tablewidth{0pt}	
\tablehead{		
\colhead{Cluster} & \colhead{[Fe/H] \tablenotemark{a}} & \colhead{$log (L/L_{\odot,
H\alpha,1})$\tablenotemark{b}} & 
\colhead{$log (L/L_{\odot, H\alpha,2})$ \tablenotemark{c}} 
& \colhead{$log (L/L_{\odot, Ca~K})$ \tablenotemark{d}} 
& \colhead{No. \tablenotemark{e}} & \colhead{$P_{1}$\tablenotemark{f}} & \colhead{$P_{2}$\tablenotemark{g}}}
\startdata
M13	 			& $-$1.54 & 1.95  & 2.79 & 1.92 & 123 & 45	& 78   \\
M15	 			& $-$2.26 & 2.36  & 2.78 & 2.36 & 110 & 22.5	& 83   \\
M92	 			& $-$2.28 & 2.05  & 2.74 & 1.96 & 64  & 18 	& 78  \\  
NGC 2808\tablenotemark{h} 	& $-$1.15 & 2.5   & 2.5  & 2.60 & 137 & 7	& 52   \\
\enddata
\tablenotetext{a}{\citet{harris01}.}	
\tablenotetext{b}{The luminosity limit of all stars showing emission in H$\alpha$.}    
\tablenotetext{c}{The luminosity limit of only RGB stars showing emission in H$\alpha$.}    
\tablenotetext{d}{The luminosity limit of all stars showing emission in \ion{Ca}{2}~K.}   
\tablenotetext{e}{Number of stars observed.}
\tablenotetext{f}{The percentage of stars from all observations showing outflow in H$\alpha$ emission wing asymmetry.}    
\tablenotetext{g}{The percentage of stars from all observations showing H$\alpha$ emission above 
$log (L/L_{\odot, H\alpha, 2})$.}
\tablenotetext{h}{Parameters of NGC~2808 for the RGB stars were
  measured by \citet{cacciari01}. No AGB stars in that
  sample are well-separated on the CMD.  Thus the RGB limit marks the
  faint limit for all stars in the cluster.}
\end{deluxetable}

\end{document}